\documentclass{aa}
\usepackage{graphicx}
\usepackage{txfonts}
\usepackage{threeparttable}
\begin{document}
  \title{870$\mu$m observations of evolved stars with LABOCA.
\thanks{Based on observations made with ESO telescopes at the La Silla Paranal Observatory under programme ID 079.F-9305A and 081.F-9320A. The reduced data can be retrieved in FITS format at CDS via anonymous ftp to cdsarc.u-strasbg.fr or via http://cdsweb.u-strasbg.fr/cgi-bin/qcat?J/A+A/}
}

  \author{
         D. Ladjal\inst{1}
         \and
         K. Justtanont \inst{2}
         \and
         M.A.T. Groenewegen\inst{3,1}
         \and
         J.A.D.L. Blommaert\inst{1}
         \and
         C. Waelkens\inst{1}
         \and
         M.J. Barlow\inst{4}
         }


  \institute{Instituut voor Sterrenkunde,
             Katholieke Universiteit Leuven, Celestijnenlaan 200D, 3001 Leuven, Belgium\\
	      \email{Djazia.Ladjal@ster.kuleuven.be}
             \and
             Chalmers University of Technology,
             Onsala Space Observatory, 
	     SE-439 92 Onsala, Sweden 
             \and
	     Royal Observatory of Belgium,
             Ringlaan 3, B-1180 Brussels, Belgium
             \and
             University College London,
             Gower Street, London,
	    WC1E 6BT, UK
	     }

  \date{Received August 2009 / accepted 4 January 2010 }

 \abstract
  {During their evolution, Asymptotic Giant Branch (AGB) stars
  experience a high mass-loss which leads to the formation of a 
  Circumstellar Envelope (CSE) of dust and gas. The mass-loss process 
  is the most important phenomenon during this evolutionary stage. 
  In order to understand it, it is important to study the physical parameters of the CSE. 
  The emission of the CSE in the (sub)millimetre range is dominated by
  the dust continuum. This means that (sub)millimetre observations
  are a key tool in tracing the dust and improving our knowledge of
  the mass-loss process.}
{The aim of this study is to use
  new sub-millimetre observations of a sample of evolved stars to
  constrain the CSE physical parameters.}
{We used aperture photometry to determine the fluxes at 870$\mu$m and to investigate
the extended emission observed using the new APEX bolometer LABoCa. 
We compute the spectral energy distribution (SEDs) using a 1D radiative transfer code, 
DUSTY which we compared to literature data. Grain properties are calculated
using both spherical grains distribution and a Continuous Distribution of Ellipsoids (CDE)
 and a comparison between the two is drawn. Synthetic surface brightness maps
have been derived from the modelling and were compared to the LABoCa
brightness maps.}
{A sample of nine evolved stars with different chemistry has been
observed with LABoCa. We detected 
the presence of extended emission around four stars.
Physical parameters of the circumstellar envelope are derived from
SED modelling, such as the dust chemical composition, the dust condensation 
temperature and the total mass of the envelope. It proves difficult
however to fit the SED and the intensity profile simultaneously. The use of the CDE leads to
"broad" SEDs when compared to spherical grains and this results in steep density distributions
($\propto r^{-2.2}$ typically).
}
  {}

  \maketitle
%

\section{Introduction}

Stars with low- and intermediate mass leave the main sequence to end their
lives on the AGB (Asymptotic Giant Branch).  This phase is characterised
by a substantial mass-loss ($> 10^{-7}\mathrm{M_{\sun}}$/yr). The
AGB-phase lasts only for about $10^6 \mathrm{yr}$ during which the star 
builds up a "circumstellar envelope" of dust and gas (CSE)
around the star.  The study of the structure of AGB envelopes is
essential for understanding the mass-loss process and the chemistry in
the different regions of the outflow.  The last decades has seen a lot
of progress in the understanding of the mass-loss mechanism and the
dust formation. This progress is directly related to the advent of
infrared astronomy and radio observations. Because the continuum
radiation in (sub)millimetre is dominated by dust emission,
this wavelength range is best suited to trace the dust emission and
by that improves our knowledge on the mass-loss process.

A number of earlier studies in the (sub)millimetre wavelength range
already provided key information about CSEs. An analytical 
method to derive the mass-loss rate of AGB stars using 
sub-millimetre data was developed by Sopka et al. (1985).
Several CSEs have been resolved and mapped (e.g., Willems \& 
de Jong 1988; Walmsley et al. 1991; Zijlstra et
al. 1992; Young et al. 1993; Groenewegen et al. 1997b; Olofsson et al. 2000). 
Various geometries have been observed (e.g., 
Knapp et al. 1998, Lopez 1999, Josselin et al. 2000) 
which indicate that the AGB phase itself
undergoes different stages which end with a strong asymmetry or a
bipolar structure precursor of the one observed in planetary
nebulae. Also, evidences pointing toward an episodic mass-loss rate
have been found (e.g., Olofsson et al 1990, 1996; Lindqvist et
al. 1996, 1999; Waters et al. 1994; Izumiura et al. 1996, 1997;
Justtanont et al. 1996, Decin et al. 2007, Dehaes et al. 2007).

In this paper, we present new sub-millimetre observations of a sample of
AGB stars observed with the APEX\footnote{This publication is based 
on data acquired with the Atacama Pathfinder Experiment (APEX). 
APEX is a collaboration between the Max-Planck-Institut f\"{u}r Radioastronomie, 
the European Southern Observatory, and the Onsala Space Observatory} 
bolometer array LABoCa at 870 $\mu$m.

The presence of extended emission around the AGB stars is investigated and the physical parameters of the CSEs are derived using SED modelling. The observations at 870\,$\mu$m proved to be not only of importance in constraining our modelling in the sub-millimeter but the observed intensity profiles could also be compared directly to the derived synthetic brightness maps.

The outline of this article is as follows: in Section~\ref{obsandreduc}, the data sample is presented and the data reduction process discussed. The total flux at 870\,$\mu$m is derived using aperture photometry and is given in Section~\ref{photometry}. The extended emission is discussed in Section~\ref{extended}. In Section~\ref{SEDmod}, the SED modelling is described and a comparison between the LABoCa maps and the synthetic brightness maps is made. The conclusions are summarised in Section~\ref{conclusion}.

\section{Observations and data reduction}
\label{obsandreduc}

The sample of stars was selected to cover various chemistry and evolutionary stages. It is composed of four Oxygen rich (O-rich) AGBs, two Carbon rich (C-rich) stars, one S-type star, one supergiant and one post-AGB star (see Table~\ref{flux}).

The observations were taken with the Large APEX Bolometer Camera
(LABoCa; Siringo et al. 2009) in service mode. 
The Atacama Pathfinder Experiment (APEX) is a 12-meter radio telescope located in Llano de Chajnantor in Chile.
The observations cover two periods in time, the first
one from 2007~July~27--August~02 and the second one from
2008~June~6--August~16. The data for CW\,Leo and VY\,CMa were
retrieved from the ESO archives and cover the period 2008~Feb~25--Feb~27 for VY\,CMa and 2008~Feb~24--Feb~29 for CW\,Leo. Despite the availability of more data for those two stars, we decided to restrict the analysis to a short period of time in order to avoid the effect of variability (for more details see Table~\ref{observations} in the appendix).

LABoCa consists of a hexagonal array of 295 channels. The half-power beam width (HPBW) is 18.6\arcsec\ and the total field of view is 11.4\arcmin. The filter passband is centred at 870~$\mu$m (345~GHz) and is 150~$\mu$m (60~GHz) wide. 
Because the array is undersampled (channel separation of 36\arcsec), special techniques are used to obtain a fully sampled map. 

For the data reduction, we used the Bolometer array Analysis Software BoA\footnote{http://www.astro.uni-bonn.de/boawiki}, which is specifically designed to handle and analyse LABoCa data.  
We followed all the steps described in the BoA User and Reference Manual (v3.1), except for the calibration part.
The standard calibration in BoA uses a list of primary calibrators (planets) and a list of secondary calibrators (well known stars calibrated using the primary calibrators).
The calibration of a science star is obtained by interpolating between the result of two calibrators (primary or secondary) bracketing the science object in time. The calibration is done to the peak flux.
While it introduces little errors when calibrating point sources, a calibration to the peak flux would overestimate the flux when used on extended sources because only a fraction of the total flux would be enclosed within the beam. In addition to the primary calibrators being slightly extended, some of the secondary calibrators are variable and also extended. In fact, two stars for which we retrieved data from the archives are used as secondary calibrators (CW\,Leo and VY\,CMa). They have been added to our analysis and will be discussed later on.
For those reasons and because we expect our science targets to show extended emission, we developed our own calibration technique using only primary calibrators and using aperture photometry. For more details about the adopted technique see Appendix~\ref{calibration}.
Additional calibration files were retrieved from the archive. For each observing night we have on average, three calibrators and three skydip measurements. 

The reduced maps are corrected for the pointing error by fitting a 2D
Gaussian to the map to determine the exact source position. 
Few maps were taken out of the analysis because of a poor signal-to-noise (S/N). The maps left were co-added to increase the S/N. For the number of scans and the total observing time for each source see Table~\ref{observations} in the appendix.
All scans were visually inspected in order to avoid the use of any corrupted data.

To derive the theoretical beam of the instrument, we used the predicted angular size for Uranus from the Astro software from the
Gildas\footnote{http://www.iram.fr/IRAMFR/GILDAS/} package to deconvolve Uranus maps to which we fitted a 2D Gaussian.
 The derived theoretical beam from 24 scans has a mean of 20.2$\pm$0.7\arcsec$\times$20.1$\pm$0.5\arcsec. The error includes the fitting error on the scans and the standard deviation for
the 24 scans. This is slightly larger than the beam size of 19.2\arcsec\ given by Schuller et al. (2009).

\section{Aperture photometry}
\label{photometry}

\subsection{Error estimation}

For the aperture photometry, we employed the IDL version of the widely
used photometry package DAOPHOT. The package proposes different
routines which allow to perform different operations on the data, such
as extract the source, fit a 2D Gaussian to your map or integrate the
flux within different apertures.

One should keep in mind that the DAOPHOT package was originally
designed to handle data from CCD cameras. 
One parameter that the {\it aper} routine requires is the conversion
from ADUs to photons, in order to estimate the photon noise. The
concept of ADUs does not apply for bolometers, but still we need to estimate
the contribution of the photon noise.

We assume that the photon rate $N_\gamma$ is:

\begin{equation}
N_\gamma= \frac{F\times S\times B}{E_\gamma}\times Q
\end{equation}

Where $F$ is the flux, $S$ the surface of the APEX
antenna, $B$ the band width of LABoCa, $E_\gamma$ is the photon energy at the 
LABoCa frequency ($ E_\gamma=h\nu $) and $Q$ the efficiency.

The only unknown of the equation is the efficiency of the instrument. 
The efficiency of bolometers is known to be close to 1 (McLean 2008). 
We verified that for any reasonable total efficiency 
the photon noise is negligible compared to the scatter in the sky level and
to the background variation. 

\subsection{Total flux estimation}
\label{totalF}

In order to estimate the total flux at the LABoCa wavelength, we selected an "ideal" aperture which would contain most of the emitted flux while keeping the background contamination low. To do so, we performed on the data the same differential aperture photometry used on the calibrators (for a complete description, see Appendix~\ref{calibration}). Essentially,  the radius of the aperture is taken to be equal to 3~$\sigma$ of the fitted Gaussian to the data, which means that 99.73\% of the total flux is within this aperture. The flux value is then corrected accordingly.

The estimated flux for each target together with the selected ideal
aperture is reported in Table~\ref{flux}. The rms of each map is also included.
The reported error on the total flux in Table~\ref{flux} is the one given by the aperture photometry routine (see DAOPHOT documentation) which includes the photon noise, the scatter in the background sky values and the uncertainty in the mean sky brightness. This error does not take into account the relative calibration uncertainty which we estimate to be $\sim$ 15\% of the total flux (see Schuller et al. 2009).

The rms values reported in Table\ref{flux} are very low compared to what is predicted by the
LABoCa observing time
calculator\footnote{http://www.apex-telescope.org/bolometer/laboca/obscalc/}
for a similar integration time and map size. This difference is due to
the calibration technique used in our case. The standard BoA
calibration encloses the total flux of the beam in the central pixel
($6\arcsec \times 6\arcsec$) when in our case the flux of the beam is
enclosed within an aperture of the size of the HPBW ($\pi
\times(\frac{HPBW}{2})^2$). This makes our rms values smaller by about 
a factor 10 than the ones derived from the maps calibrated following the
standard BoA calibration.

The derived total flux at 870$\mu$m is included in the SED modelling 
and is in fair agreement with previous sub-millimetre observations given the intrinsic 
variability of some of the stars (see Fig.~\ref{ocet} to Fig.~\ref{figLast} in the appendix).

\begin{table*}
\centering
\caption{Measured flux at 870~$\mu$m.}
\label{flux}
\begin{tabular}{llllrrrrr}
\hline
IRAS name & Alternative ID & Spectra type & Variability & Flux at 870~$\mu$m & Error$^a$ & rms & Ideal aperture$^b$ \\ 
& & & & [mJy] & [mJy] & [mJy/pixel] & [\arcsec]  \\
\hline
01037$+$1219 & WX\,Psc & M9 & LPV & 145 & 3 & 0.38 & 42 \\
02168$-$0312 & $o$\,Cet & M7 & Mira & 339 & 3 & 0.51 & 40  \\
07209$-$2540 & VY\,CMa & M3/M4II &   & 1304 & 15 & 2.98 & 35  \\
09452$+$1330 & CW\,Leo & C & & 7243 & 9 & 1.98 & 48 \\
16342$-$3814 &  & PAGB & & 602 & 3 & 0.56 & 37 \\
17411$-$3154 & AFGL\,5379 & OH/IR & & 304 & 4 & 0.64 & 39  \\ 
19244$+$1115 & IRC\,$+$10420 & F8Ia & & 243 & 3 & 0.46 & 38  \\
22196$-$4612 & $\pi$\,Gru & S & SRb & 143 & 3 & 0.38 & 42  \\
23166$+$1655 & AFGL\,3068 & C & Mira & 299 & 5 & 0.85 & 38 \\ 
\hline
\end{tabular}

\flushleft{\footnotesize{(a) Internal error. The absolute calibration error is about 15\% of the total flux.}}
\flushleft{\footnotesize{(b) Corresponds to 3~$\sigma$ of the 2D gaussian fit to the data (see section~\ref{totalF}).}}
\end{table*}

\section{The extended emission}
\label{extended}

To investigate the presence of any extended emission at 870~$\mu$m,
we adjusted an elliptical Gaussian to the surface brightness map of
each target with a free orientation of the axes of the ellipse
(relevant in the case of asymmetric sources). The FWHM of the 
ellipse in both directions is compared to the FWHM of the theoretical beam. A source is considered extended if the difference in size between the
source and the theoretical beam combining the errors is
larger than a factor 3 in at least one direction (see parameter $s$ in
Table~\ref{2Dfit}). A position angle is derived for the resolved sources using a least square minimisation and is given in Table~\ref{2Dfit} (see parameter $PA$). We compute for the resolved sources both the major axis $a$ and the minor axis $b$ of the ellipse
(see Table~\ref{2Dfit}) and use the new values to investigate the
extended emission.

From Table~\ref{2Dfit}, extended emission at 870 $\mu$m is seen
only for CW\,Leo and $\pi$\,Gru in both directions and WX\,Psc and
$o$\,Cet in one direction.

Young et al. (1993) detected an extended 
emission at 60$\mu$m for CW\,Leo, $\pi$\,Gru and $o$\,Cet but did not observe 
any extended emission for VY\,CMa, AFGL\,3068, IRC\,+10420, WX\,Psc and AFGL\,5379.

A 250\arcsec\ $\times$ 250\arcsec\ cut of the LABoCa maps  for the resolved sources can be seen from Fig.~\ref{CW_cont} to Fig.~\ref{OMI_cont}. A 3$\sigma$ level above the noise contour is drawn. The FWHM of the adjusted ellipse in both $x$ and $y$ directions can be found in Table~\ref{2Dfit}.

The four sources for which extended emission is found at 870\,$\mu$m will be discussed in the following subsections.

\subsection{CW\,Leo}

CW\,Leo shows the largest extended emission of our set with a
25\arcsec\ FWHM spherically symmetric CSE at 870~$\mu$m (see Fig.~\ref{CW_cont} and Table~\ref{2Dfit}). 
This is not surprising, as this nearby carbon star is subject to a high mass loss rate and is surrounded by a thick CSE. The circumstellar shell has been imaged at different wavelengths, 
in the visible by Mauron and Huggins (2000), in the NIR by Le\~{a}o et al. (2006), 
at 1.3~mm by Groenewegen et al. (1997b) and using CO lines by Huggins et al. (1998) and Fong et al. (2004).

CW\,Leo is known for being variable in the K band with a magnitude variation from 1.9 to 2.9 mag and in the M band with a variability $\sim$ 0.6 mag, the estimated period is $\sim$ 638 days (Dyck et al. 1991). It is also variable at radio-wavelengths (1.3~cm) with a 25\% flux variation with a period of 535$\pm$52 days (Menten et al. 2006). This implies that CW\,Leo
probably has variability at 870 $\mu$m. Despite the brightness of CW\,Leo, the variability and the extended emission at the LABoCa wavelength makes it unsuitable as a flux calibrator.

\subsection{$\pi$\,Gru}
\label{pigru}

The 2D Gaussian fit of the LABoCa map of $\pi$\,Gru shows an extended emission with a FWHM of 23.3\arcsec\ following the east-west direction and 22\arcsec\ following the north-south direction (see Fig.~\ref{PI_cont} and Table~\ref{2Dfit}). Beyond the FWHM, $\pi$\,Gru appears even more elliptical in the LABoCa map with a structure of about 60\arcsec\ by 40\arcsec.

An asymmetry in the CO line profiles of this star has been observed by Sahai (1992) and by Knapp et al. (1999) with two velocity components being derived. A fast outflow with a speed greater than 38~km$s^{-1}$ and perhaps as high as 90~kms$^{-1}$  and a "normal" outflow with a speed of about 11~kms$^{-1}$. Sahai (1992) suggested that the asymmetry in the line profiles is the result of a fast bipolar outflow collimated by an equatorial disk (major axis direction east-west). Knapp et al. (1999) proposed a model with a disk to explain the observed CO line profiles.  In their model, the disk is tilted by 55$^\circ$ to the line of sight with a major axis lying east-west. The asymmetry in the CO line profiles would then be related to the northern and southern halves of the disk, the fast outflow was not included in the modelling.

According to Huggins (2007), the creation of a disk/jet structure in AGB stars needs a binary system. In fact, $\pi$\,Gru is the primary of a wide binary system with a G0V star as secondary (Proust et al. 1981; Ake \& Johnston 1992). The large separation (of about 2.71\arcsec) makes it difficult for the companion to be responsible for the bipolar outflow. One possible explanation given by Chiu et al. (2006) is the presence of a much closer companion that has escaped detection and which would be the source of the high-velocity outflow. This possibility is supported by Makarov \& Kaplan (2005) and Frankowski et al. (2007) who found significant discrepancies between Hipparcos and Tycho-2 proper motions of the star which indicates additional orbital motion with an orbital period shorter than the 6000 years derived by Knapp et al. (1999).

Sacuto et al. (2008) attempted to resolve the disk using MIDI/VLTI interferometric data at 10$\mu$m but the data were not conclusive. They do detect the presence of an optically thin CSE but it does not depart from a spherical shape. Sacuto et al. (2008) argue that the disk can be confused for a spherical shell if the system opening angle is larger than the interferometric angular coverage (60$^\circ$). A poor $uv$ coverage can also be a reason for not resolving the disk.

If we assume that the extended emission observed in the LABoCa map is from the disk, geometrical constraints and the derived major and minor axes would imply a disk tilted by about 70$^\circ$ to the line of sight with a major axis lying east-west, in agreement with Knapp et al. (1999). This higher angle could also explain why it is easy to confuse the disk for a spherical CSE.

\subsection{WX\,Psc}
\label{wxpsc}

WX\,Psc is an O-rich star known for being variable at infrared
wavelengths with a magnitude variation from 1.29 magnitude in the
M-band up to 2.83 magnitudes in the J-band, with a period of 660 days
(Le Bertre 1993), $\Delta m=$1.5 at 10$\mu$m (Harvey et al. 1974) and
$\Delta m=$1.2 at 18~cm (Herman \& Habing 1985). 

The LABoCa data show an extended emission of 23.6\arcsec\ following
one direction with a position angle of 84$^\circ$ (see Fig.~\ref{WX_cont} and Table~\ref{2Dfit}).

From previous CO interferometric observations, Neri et al. (1998) suggest that the star
has two CSE components, a spherically symmetric component with a radius of
29.6\arcsec\ and an elliptical component with a major axis of
9.8\arcsec\ and a minor axis of 6.8\arcsec\ with a position angle of
$-$45$^\circ$.

Hofmann et al. (2001) conducted interferometric observations of WX\,Psc in the J-, H- and K-bands. In the J-band, the results show a clear elongation along a symmetry axis with a position angle of $-$28$^\circ$. This asymmetric structure is composed of two components, a compact elliptical core with a major axis of 154~mas and a minor axis of 123~mas and a fainter "fan" shape structure with the opening angle of the fan from $-$8$^\circ$ to $-$48$^\circ$, out to distances of $\sim$200~mas. This "fan" like structure is hardly seen in the H and the K band, where the dust shell displays a spherical symmetry. The J-band structure has been modelled by Vinkovi\'{c} et al. (2004) who could reproduce the same structure by constructing a 2D radiative transfer model considering a bipolar jet which would sweep up material in the slower spherical wind and create a cone as seen in the NIR data. Inomata (2007) claimed the existence of such a bipolar outflow after analysing H$_2$O maser spectra of WX\,Psc and deriving a structure position angle in agreement with what found by Hofmann et al. (2001).

A V-band image of WX\,Psc, obtained with the VLT by Mauron \& Huggins
(2006), shows a spherically symmetric shell out to
$\sim$50\arcsec\ distance while in an HST-ACS image at 816~nm a highly
asymmetric core is seen with an extension out to
$\sim$~0.4\arcsec\ with a $-$45$^\circ$ position angle. Mauron
\& Huggins conclude that WX\,Psc is similar to CW\,Leo in respect to
the presence of a circular symmetry at large scale together with a
strong asymmetry close to the star.

The asymmetric extended emission observed in the LABoCa map of WX\,Psc disagrees with Mauron
\& Huggins (2006) assumption by showing that the asymmetry in the dust structure can still be observed at bigger scales (up to 40\arcsec). This asymmetric dust structure is also in disagreement with the spherical gas envelope observed at 1.3~mm by Neri et al. (1998). This could be the result of an extremely complex mass-loss history with sudden changes in the mass loss rate during the life span of the star resulting in multiple shells with different geometries.

\subsection{$o$\,Cet}
\label{omicet}

Also known as Mira, it is the prototype for Mira-type variables. It is
also one of the closest O-rich AGB stars with an estimated distance of
131$\pm$18~pc (Perryman et al. 1997). Its variability reaches 3
magnitudes in the V-band with a period of 331 days (GCVS).

The LABoCa map shows an asymmetric CSE with an elongated core of 22.8\arcsec\ with a position angle of 79$^\circ$ (see Fig.~\ref{OMI_cont} and Table~\ref{2Dfit}).

From previous studies, we know that Mira shows asymmetry on all scales. At large scale a turbulent wake 2$^\circ$ long was observed in emission in the ultraviolet (Martin et al. 2007) and in the 21~cm HI line (Matthews et al. 2008). 
At small scale the star itself is asymmetric (Karovska et al. 1997). 

The data of Josselin et al. (2000) show an asymmetric "peanut" like molecular envelope lying roughly east-west with two "holes" located at about 4\arcsec\ north and south with three velocity components which suggest the presence of a bipolar outflow. Unlike $\pi$\,Gru however, the bipolar outflow velocity is low and close to the velocity of the spherical component which is about 8~kms$^{-1}$ and as such cannot account for the observed asymmetry. In order to explain the unusual shape of Mira, Josselin et al. (2000) suggest that the departure from sphericity occurs very early on the AGB, if not since the beginning. Furthermore, the bipolar outflow could accelerate as the star evolves along the AGB phase.

Like $\pi$\,Gru, $o$\,Cet is part of a binary system (Joy 1926; Karovska \& Nisenson 1993) with a close companion at a distance of 0.61\arcsec$\pm$0.03\arcsec\ and a position angle of 111$^\circ$. The companion is expected to be a relatively hot star (a main sequence star or a white dwarf) (Karovska \& Nisenson 1993). 

Using Spitzer data, Ueta (2008) shows the presence of an astropause (a stellar analogue of the heliopause) 
 at 160~$\mu$m with a slightly extended core at a position angle of 70$^\circ$ and derives a correlation between the dust emission at 160~$\mu$m and the turbulent wake observed in the ultraviolet. The position angle of 79$^\circ$ that we find at 870$\mu m$ is close to the 70$^\circ$ position angle found by Ueta (2008) at 160$\mu m$ and implies that the extended ellipsoid lies east-west, which is also in agreement with the findings of Josselin et al. (2000).

\begin{table*}
\centering
\caption{Results of the 2D elliptic Gaussian fit.}
\label{2Dfit}
\begin{tabular}{llccrcccc}
\hline
IRAS name & Alternative ID & FWHM$_x$ & FWHM$_y$ & $PA$ & $a$ & $b$ & $s_x$ & $s_y$\\ 
& & [\arcsec]  & [\arcsec] & [$^\circ$] & [\arcsec]  & [\arcsec] & & \\
\hline
01037$+$1219 & WX\,Psc & 23.6$\pm$0.4 & 21.7$\pm$0.4 & 84 & 23.6 & 21.7 & 4.2 & 2.5 \\
02168$-$0312 & $o$\,Cet & 22.7$\pm$0.2 & 21.4$\pm$0.2 & 79 & 22.8 & 21.4 & 3.6 & 2.4 \\
07209$-$2540 & VY\,CMa & 19.2$\pm$0.2 & 20.0$\pm$0.3 & & & & 1.4 & 0.2  \\
09452$+$1330 & CW\,Leo & 25.0$\pm$0.1 & 25.2$\pm$0.1 & & & & 6.8 & 10.0 \\
16342$-$3814 &  & 20.7$\pm$0.1 & 20.4$\pm$0.1 & & & & 0.7 & 0.6 \\
17411$-$3154 & AFGL\,5379 & 20.9$\pm$0.3 & 21.2$\pm$0.3 & & & & 0.8 & 1.9 \\ 
19244$+$1115 & IRC\,$+$10420 & 22.0$\pm$0.2 & 20.9$\pm$0.2 & & 22.0 & 20.9 & 2.5 & 1.5 \\
22196$-$4612 & $\pi$\,Gru & 23.1$\pm$0.4 & 22.2$\pm$0.4 & -68 & 23.3 & 22.0 & 3.6 & 3.3 \\
23166$+$1655 & AFGL\,3068 & 21.2$\pm$0.4 & 20.7$\pm$0.3 & & & & 1.2 & 1.0 \\ 
\hline
Uranus & & 20.4$\pm$0.7 & 20.3$\pm$0.5 & & & & & \\
PSF & & 20.2$\pm$0.7 & 20.1$\pm$0.5 & & & & & \\
\hline\hline
\end{tabular}
\flushleft{\footnotesize{\textit{Note to Table~\ref{2Dfit}}: $PA$ is the position angle of the major axis of the ellipse, $a$ and $b$ the major and minor axis, $s_x$ and $s_y$ represent the relative size of the source to the beam following $x$ and $y$ (see section~\ref{extended}).}}
\end{table*}

\section{SED Modelling and synthetic brightness maps}
\label{SEDmod}

\subsection{The radiative transfer code DUSTY}
\label{dusty}

Theoretical spectral energy distributions (SEDs) were computed using the 1D radiative transfer code
DUSTY\footnote{http://www.pa.uky.edu/~moshe/dusty/} (Ivezi\'{c} et al. 1999).

Originally, DUSTY considers only spherical dust grains, but we added a
subroutine to the main code to also be able to consider a Continuous
Distribution of Ellipsoids (CDE) as the distribution of grain shapes. The ellipsoids are randomly oriented and have different shapes but have always the same volume. CDE has been very successful in reproducing the SED features of C-rich stars
(e.g. Hony et al. 2002) and the observed features of silicates (e.g. Bouwman et al. 2001).

\begin{figure}
\vspace{0cm}
\hspace{0cm}
\resizebox{8cm}{!}{ \includegraphics[angle=90]{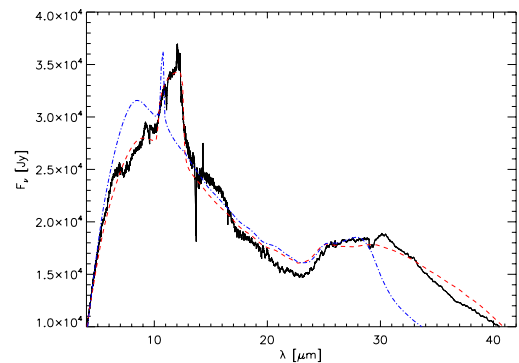}}
\caption{Comparison between the shape of the dust features of the best model for ISO-SWS data of CW\,Leo (black full line) using spherical grains (blue dash-dot line, 2\% SiC, 4\% MgS and 94\% amorphous carbon) and using CDE (red dashed line, 12\% SiC, 22\% MgS and 66\% amorphous carbon). The SiC feature is at 11$\mu$m and the MgS feature is around 30$\mu$m.}
\label{sphere_CDE}
\end{figure}

In Fig.~\ref{sphere_CDE} the difference in shape of the dust features
(in this case SiC at 11$\mu$m and MgS around 30$\mu$m) when using a
spherical grain distribution or CDE is shown.
The ISO-SWS/LWS spectra of CW\,Leo is also plotted in order to compare the shape of the modelled features to the observed data. When using the CDE, the features are broader and are a better match to the observed data. Additionally, the position of the peaks are shifted compared to the spherical case and in better agreement with the observations.

\begin{figure}
\vspace{0cm}
\hspace{0cm}
\resizebox{8cm}{!}{ \includegraphics[angle=90]{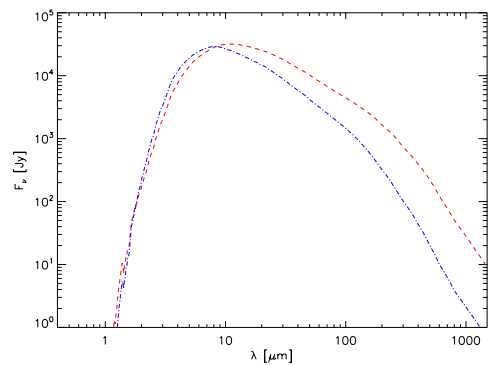}}
\caption{Comparison between spherical grains (blue dash-dot line, 100\% amorphous carbon) and CDE (red dashed line, 100\% amorphous carbon) for a same density distribution and optical depth.}
\label{sphere_CDE_density}
\end{figure}

One of the consequences in using the CDE is that for a same density distribution we obtain a much higher level of flux in the FIR than when using spherical grains (see Fig.~\ref{sphere_CDE_density}). This leads to the use of steeper density distributions in order to fit the observed data in that wavelength range.

We adopted the CDE distribution for all the models.
For a more detailed description of the CDE and for the scattering and absorption coefficient equations, see Min et al. (2003).

\subsection{The dust chemical composition}

DUSTY comes with a built-in library of optical properties for six
common dust components but it can also handle additional components if
provided with the adequate optical properties.

For O-rich stars we used the recent
MARCS\footnote{http://marcs.astro.uu.se/} stellar atmosphere models for cool stars (see
Gustafsson et al. 2008) as a radiative source. The models cover a wavelength range from 0.2$\mu$m to 20$\mu$m. Beyond 20$\mu$m we used Rayleigh$-$Jeans approximation. 

Concerning the dust composition, we investigated the following dust species:

- Amorphous silicate (Mg$_x$Fe$_{2-x}$SiO$_4$): it is the dominant component in O-rich stars. Only by using amorphous silicate we can reproduce the overall shape of the SED. It is also responsible for the prominent 9.7~$\mu$m feature and the 18~$\mu$m feature. There are different sets of silicates which differ slightly in shape and in the position of the features. For this work we investigated three different types of silicates: Draine \& Lee (1984), Ossenkopf et al. (1992) and Dorschner et al. (1995). In some cases Dorschner et al. (1995) grains give a better agreement with the observed silicate features where in some other cases Ossenkopf et al. (1992) grains are a better representation. Ossenkopf et al. (1992) grains are considered better suited for cool oxygen-rich dust which are representative in OH/IR stars (see Ossenkopf et al. 1992 for a discussion).

- Amorphous aluminium oxide (Al$_2$O$_3$): In the dust condensation
sequence, it is considered as the first condensate. It was
successfully used to reproduce the broad 12~$\mu$m feature in IRAS
spectra of AGB stars (Onaka et al. 1989 ; Egan \& Sloan 2001), UKIRT and ISO-SWS spectra
(e.g., Speck et al. 2000 ; Maldoni et al. 2005 ; Blommaert et al. 2006). 
For this analysis we are using the grains optical properties by Begemann et al. (1997).

- Metallic iron (Fe): it was successfully used by Kemper et al. (2002)
to correct the overestimated opacity in the near-infrared (NIR)
for OH/IR stars.

\vspace{6mm}

For C-rich stars we used the synthetic models of Loidl et al. (2001) and investigated 3 carbon-rich dust species:

- Amorphous carbon (AmC here) : It is the main component in C-rich stars. Two different sets of amorphous carbon grains were tested ; Preibisch et al. (1993) and Rouleau \& Martin (1991). The absorption coefficient of the two components is very similar in NIR and mid-infrared but becomes different in the far-infrared (FIR) where Rouleau \& Martin (1991) grains give more flux. Using a CDE distribution, Preibisch et al. (1993) grains give a better fit to the data in the sub-millimetre range than Rouleau \& Martin (1991). In the following, AmC will refer only to Preibisch et al. (1993) grains.

- Magnesium sulfide (Mg$_{0.9}$Fe$_{0.1}$S): It has been used by Hony \& Bouwman (2004) to reproduce the 30~$\mu$m feature. In this work we use the optical properties of Preibisch et al. (1993).

- Silicon carbide (SiC): Two different types of silicon carbide were
tested, P\'{e}gouri\'{e} (1988) and Pitman et al. (2008) grains. The
latter turned out to be a better candidate for modelling the 11~$\mu$m
feature, regarding both shape and peak position. The peak position was
especially an issue as P\'{e}gouri\'{e} (1988) grains showed an
important offset to the blue compared to the observed data.
In the following sections we will only refer to the Pitman et al. (2008) grains.

\subsection{Modelling}

A grid of models was constructed by varying the following physical parameters : the effective temperature
$T_{\rm eff}$ in the range of 2400~K to 3500~K), the dust condensation temperature $T_{\rm c}$ ranging from 500~K to 1200~K, the dust chemical composition, the optical depth $\mathbf{\tau}_{\lambda}$ at 0.55~$\mu$m and the density power index which was varied from $-$2.7 to $-$1.7.  The maximum outer radius of the shell was set to be 10,000 times the inner shell radius and the dust grain volume when averaged by a sphere is equal to $4/3 \pi a^3$ with $a$ set to 0.05 $\mu$m. More details about the construction of the grid of models can be found in Appendix~\ref{grid}.

Comparisons between the models and the data were made using an automated fitting routine and the best fit was selected based on a least square minimisation technique. 
The models were compared to the ISO Short Wavelength Spectra (SWS), retrieved from Sloan
et al. (2003) database, ISO Long Wavelength Spectra (LWS) spectra retrieved from ISO archives
and when available, photometric data from the visible to the millimetre were retrieved from the literature (see Table~\ref{photodata} for references). The LWS spectrum was shifted to match the SWS flux level. SWS and LWS spectra were available for all the stars except for IRC\,+10420 for which we have only the SWS spectrum. 
The consistency of the flux calibration between the SED and the IRAS photometry was checked for all the stars. The ISO data of WX\,Psc shows a significant discrepancy compared to the IRAS flux, so we decided to recalibrate the SED to the IRAS photometry. The SED was corrected by a factor of 0.6. Hofmann et al. (2001) also suggest the recalibration of the ISO data of WX\,Psc to the IRAS photometry.
The spectroscopic data and the photometry points were weighted by taking into account their respective spectral resolution. Since one of the main goals of this study is to compare the synthetic brightness maps at 870$\mu$m to LABoCa maps, we enhanced the weight of the LABoCa point compared to the other sub-millimetre measurements to ensure a good correspondence.

\begin{table*}
\scriptsize
\begin{threeparttable}[c]
\setlength{\tabcolsep}{1.5mm}
\centering
\caption{SED fitting results.}
\label{SEDfit}
\begin{tabular}{lcccccccccccccc}
\hline
O-rich stars  \\
\hline
 Name & $T_{\rm eff}$ & $T_{\rm c}$ & $\mathbf{\tau} _{\lambda}$ & n & $D$ & $V_{gas}$ & $L$ (x10$^{4}$) &$\dot{M}$ (x10$^{-6}$) & $M_{\rm d}$ (x10$^{-5}$)& ${M}$ (x10$^{-3}$) & Am. Silicate & Alumina & Fe & $\dot{M}_{ref}$ (x10$^{-6}$) \\ 
             &  [K]          & [K]           &          &      & [Kpc] &[Kms$^{-1}$]&L$_{\sun}$&[M$_{\sun}/yr$]& [M$_{\sun}$] & [M$_{\sun}$] & [\%]  	      &[\%]		&[\%]	 & [M$_{\sun}/yr$]  \\
\hline
 WX\,Psc & 2500 & 1010 & 9 & -2.0 & 0.75\tnote{a} & 20\tnote{b} & 1.1& 13 & 17 & 36  & 98 Ossen. &  & 02 & 0.3$-$8\tnote{b}  \\
 $o$\,Cet & 2600  & 1000 & 0.12 & -1.9 & 0.128\tnote{c} & 8\tnote{b} &1.4& 0.11  & 0.11 & 0.11 & 100 Dorsch. &  & & 0.5\tnote{k}  \\
 VY\,CMa & 2500 & 1100 & 13 & -2.1 & 1.14\tnote{r} & 47\tnote{b} &27.5& 203 & 600 & (...) & 98 Ossen. &  & 02 & 100\tnote{n}$-$300\tnote{o}  \\
 IRAS 16342$-$3814 & 2600 & 1000 & 480 &-2.2 & 2\tnote{i} &46\tnote{h} &0.04& 319 & 128 & \textless\ 140 & 75 Ossen. &  & 25 &300\tnote{i} \\
 AFGL\,5379 & 2600 & 700 & 52 &-2.3 & 1.19\tnote{d} & 24\tnote{b} &0.5& 46 & 97 & \textless\ 250 & 97 Dorsch. &  & 03 & 9$-$200\tnote{q}\\ 
 IRC\,$+$10420 (ISO) & 7000 & 750 & 5 & -2.2 &5\tnote{e} & 42\tnote{b} &47& 693 & 1980 & \textless\ 4400 & 99.5 Dorsch. &  & 0.5 & 700$-$900\tnote{m}  \\
 IRC\,$+$10420 (LRS)     & 7000 & 750 & 4  & -2.2 & 5\tnote{e} &  42\tnote{b}&39 & 260 & 522 &\textless\ 1600 & 90 Dorsch. & 08 & 02 & 700$-$900\tnote{m} \\
 $\pi$\,Gru & 3000 & 900 & 0.04 & -2.0 &0.153\tnote{c} & 11\tnote{g} &1.04& 0.05 & 0.01 & 0.06 & 50 Ossen. & 50 &  & 0.2$-$0.7\tnote{l}  \\
\hline\hline
C-rich stars  \\
\hline
Name & $T_{\rm eff}$ & $T_{\rm c}$ & $\mathbf{\tau} _{\lambda}$ & n & $D$ & $V_{gas}$ & $L$ (x10$^{4}$)&  $\dot{M}$(x10$^{-6}$) & $M_{\rm d}$ (x10$^{-5}$)& ${M}$ (x10$^{-3}$) & Am. C & SiC & MgS & $\dot{M}_{ref}$(x10$^{-6}$) \\ 
            &  [K]          & [K]           &          &      & [Kpc] &[Kms$^{-1}$]&L$_{\sun}$&[M$_{\sun}$]&[M$_{\sun}$] & [M$_{\sun}$] & [\%]  	      &[\%]		&[\%]	 & [M$_{\sun}/yr$]  \\
\hline
 CW\,Leo & 2650 & 1050 & 25 & -2.3 &0.135\tnote{j} & 17\tnote{f} &0.9& 7 & 1.2 & 6.2 & 66 & 12 & 22 & 22\tnote{j} \\
 AFGL\,3068 & 2650  & 1000  & 135 & -2.5 &1.05\tnote{d} & 16\tnote{f} &1.4& 110 & 3.6 & \textless\ 300 & 78 & 04 & 18 & 20$-$62\tnote{p} \\
\hline\hline
\end{tabular}
\begin{tablenotes}[para]
References.
	(a) van Langevelde et al. (1990);
	(b) Kemper et al. (2003);
	(c) Hipparcos (Perryman \& ESA 1997);
	(d) Yuasa et al. (1999);
	(e) Jones et al. (1993);
	(f) Groenewegen et al. (1996);
	(g) Knapp et al. (1999);
	(h) He et al. (2008);
	(i) Sahai et al. (1999);
	(j) Groenewegen et al. (1998b);
	(k) Loup et al. (1993);
	(l) Groenewegen et al. (1998a);
	(m) Dinh-V-Trung et al. (2009);
	(n) Jura \& Kleinmann (1990);
	(o) Harwit et al. (2001);
	(p) Ramstedt et al. (2008);
	(q) Kastner (1992);
	(r) Choi et al. (2008).

\end{tablenotes}
\end{threeparttable} 
\flushleft{\scriptsize{\textit{Note to Table~\ref{SEDfit}}: $T_{\rm eff}$ is the effective temperature, $T_{\rm c}$ the dust condensation temperature, $\mathbf{\tau} _{\lambda}$ is the optical depth at 0.55$\mu$m, n is the density power index, $D$ the distance to the star,  $V_{gas}$ the terminal velocity derived from CO line measurements, $L$ the luminosity, $\dot{M}$ the mass-loss rate, $M_{\rm d}$ is the dust mass , ${M}$ the total dust and gas mass-loss, $\dot{M}_{ref}$ the mass-loss derived from previous studies.}}
\end{table*}

\subsubsection{SED fitting results}

The modelling results are summarised in Table~\ref{SEDfit}. Fig.~\ref{ocet} to Fig.~\ref{figLast} show the comparison between the models and the data.  We can see that the synthetic models are in most cases in good agreement with the observed data despite the adopted assumption of spherical symmetry which is not valid for all the stars (see Section~\ref{extended}).

Concerning $\pi$\,Gru, the discrepancy between the model and the real
SED may be explained by the fact that the O-rich MARCS model used as
radiative source is not adapted for S-stars.
As the CSE is optically thin, the features seen in the synthetic model are mostly
that of the MARCS model.

The model for IRC\,+10420 agrees poorly with the observations between 15$\mu$m and 100$\mu$m. The ISO-SWS spectrum shows a high flux level between 15$\mu$m and 45$\mu$m. This
high level of flux could not be reproduced by our modelling (see Fig.~\ref{fig10420}). This "jump" is not seen in the IRAS Low Resolution Spectrum (LRS) what suggests that the "jump" may be an artefact of the SWS data reduction. Connecting together the different SWS segments can be difficult as each segment relies on the calibration of the previous one, this issue is discussed in Sloan et al. (2003).
Previous analysis of IRC\,+10420 (e.g. Oudmaijer et al. 1996) used the LRS spectrum in the SED fitting of the star which we could easily model in Fig.~\ref{fig10420}.

In the case of the OH/IR star AFGL\,5379, the FIR excess is not well
reproduced by the modelling (see Fig.~\ref{fig5379}) despite the use
of metallic iron which was suggested by Kemper et al. (2002) to
correct the FIR excess of OH/IR stars.

From the output of DUSTY, it is deduced that the photospheric flux of the central source is negligible at 870$\mu$m in the case of most of our stars except for $o$\,Cet and $\pi$\,Gru for which the photospheric flux contributes up to 50\% to the total flux observed at 870$\mu$m.

\subsubsection{Density distribution}

Contrary to previous analyses which used a density distribution following an $r^{-2}$ density power law or even shallower densities to match the observations (e.g. Sopka et al. 1985, Groenewegen 1997a ; Blanco et al. 1998 ; Lorenz-Martins et al. 2001 ; Gautschy-Loidl et al. 2004), our models follow a steeper density distribution ( $\propto r^{-2.3}$ for AFGL\,5379 and CW\,Leo, $\propto r^{-2.5}$ for AFGL\,3068) (see the density power index "n" in Table~\ref{SEDfit}).

The main difference between those analyses, with a shallow density distribution, and our analysis is that they used spherical grains distribution while we used CDE. Section~\ref{dusty} discusses what the CDE does to the shape of dust features compared to the spheres distribution. We clearly see in the spectral energy distribution in Fig.~\ref{sphere_CDE} that CDE gives more realistic shape and peak position of the dust features (see also Min et al. 2003) but at the same time CDE gives a higher level of flux in the sub-millimetre (see Fig.~\ref{sphere_CDE_density}). This leads to the use of a steeper density distribution law than $r^{-2}$ in order to reproduce the observed data with the uncertainty on the density power being $\pm 0.05$.

\subsubsection{Mass$-$loss}

The derived physical parameters from the SED modelling can be used to constrain 
the mass-loss rate $\dot{M}$ of our sample of stars.

For a constant mass-loss rate and a constant dust velocity, the optical depth $\mathbf{\tau} _{\lambda}$ is related to the mass-loss rate via (Groenewegen et al. 1998b)\,: 

\begin{equation}
\label{massloss}
\mathbf{\tau} _{\lambda} = 5.405\times10^{8}\frac{\dot{M}\ \Psi\ Q_{\lambda}/{a}}{r_{\rm c}\ R_*\ V_{\rm d}\ \rho_{\rm d}}
\end{equation}

Where $\dot{M}$ is in M$_{\sun}$/yr, $V_{\rm d}$ is the dust velocity in km s$^{-1}$ , $R_*$ is the stellar radius in solar radii, $r_c$ is the inner dust radius in stellar radii, $Q_{\lambda}$ is the absorption coefficient, $a$ is the dust grain radius in cm which in the case of CDE is the radius of a sphere with the same volume as the ellipsoid, $\Psi$ is the dust-to-gas mass ratio which we assumed to be 0.005 and $\rho_{\rm d}$ is the grain density in g cm$^{-3}$.

The parameters that are derived from the SED modelling are the optical depth at a specific wavelength $\mathbf{\tau}_{\lambda}$, $R_*$ and $r_{\rm c}$ which can be computed using the effective temperature $T_{\rm eff}$ and the dust condensation temperature $T_{\rm c}$ and by assuming a distance. The assumed distances are listed in Table~\ref{SEDfit} together with the references.

We assumed the dust velocity $V_d$ as being equal to the gas terminal velocity $V_{gas}$ and used expansion velocities derived from CO line measurements (see Table~\ref{SEDfit}). From equation~\ref{massloss} we can see that a change on the dust velocity has a linear affect on the mass-loss rate.

The dust grain density for silicates varies from 2.8 g cm$^{-3}$ to 3.3 g cm$^{-3}$ (see Ossenkopf 1992). We assumed an average value of 3 g cm$^{-3}$ for all O-rich stars. Concerning C-rich stars, we adopted the density distribution of the amorphous carbon used in our modelling, which is 1.85 g cm$^{-3}$ (see Preibisch et al. 1993 and Bussoletti et al. 1987).

The mass absorption coefficient $\kappa_{\lambda}= \frac{3Q_{\lambda}}{4a\rho}$ follows directly from DUSTY outputs and ranges from $\kappa_{870\mu m} \sim$ 2\,cm$^{2}$g$^{-1}$ for M-stars and $\kappa_{870\mu m} \sim$ 35\,cm$^{2}$g$^{-1}$ for C-stars with intermediate values of 4\,cm$^{2}$g$^{-1}$ and 8\,cm$^{2}$g$^{-1}$ for the OH/IR star AFGL\,5379 and the S-star $\pi$\,Gru, respectively. For the post-AGB star IRAS\,16342$-$3814, we find $\kappa_{870\mu m} \sim$ 20\,cm$^{2}$g$^{-1}$ because of the the presence of a high fraction of iron which has a high opacity. These results are a magnitude higher than what is predicted by Draine (1981) for grains in the interstellar medium. Sopka et al. (1985) claim an order of magnitude uncertainty on the parameter when considering circumstellar grains rather than interstellar.

The derived mass-loss rates are of the same order as the one estimated in previous studies (see Table~\ref{SEDfit}). The difference is within the predicted uncertainty for $\dot{M}$ estimates. One should keep in mind the dependence of $\dot{M}$ on the distance which can be very uncertain for AGB stars and also the dependence on the gas-to-dust ratio which is very difficult to constrain and can highly vary from one star to another. Ramstedt et al. (2008) discuss the reliability of mass-loss rate estimates for AGB stars and compare $\dot{M}$ values from CO line measurements and from SED modelling. The agreement between the two methods is within a factor $\sim$3  which they consider as the minimal uncertainty in present mass-loss rate estimates. 

By assuming a constant mass-loss and a constant dust velocity and by using the derived size of each source, we can estimate the total mass of the CSE for each star. The minimum diameter of the CSE is assumed to be 3 times the average FWHM after deconvolution from the beam. The derived total mass of the CSE $M$ is listed in Table~\ref{SEDfit}. In the case of unresolved CSEs, the calculated total mass represents an upper limit as the size derived from the observations is overestimated.

\subsubsection{Dust mass in CSEs}

Using a simple approach, we can estimate the total dust mass in the
envelope M$_{\rm d}$, assuming the dust at 870${\mu}$m is optically thin, by (Hildebrand, 1983) :
\begin{equation}
M_{\rm d} = 9.52\times 10^{36} \frac{F_{\lambda} D^{2}}{\kappa_{\lambda}B_{\lambda}(T_{\rm d})}
\end{equation}
Where $F_{\lambda}$ is the observed LaBoCa flux in erg~s$^{-1}$cm$^{-3}$ (corrected from photospheric contamination for $\pi$\,Gru and $o$\,Cet), D is the distance in pc,
$\kappa_{\lambda}$ is the dust opacity in cm$^{2}$g$^{-1}$ and $B_{\lambda}(T_{\rm d})$
is the Planck function at the temperature where the dust emission peaks in erg~s$^{-1}$cm$^{-3}$.
From the parameters in Table~\ref{SEDfit} and the DUSTY outputs, we can determine
the total dust mass for each of our object. The main uncertainty here is
the determination of the dust temperature, the distance and the mass absorption coefficient. However, this approach should give us an order-of-magnitude estimate of the dust mass (see M$_{\rm d}$ in Table~\ref{SEDfit}) in the envelope.

\subsection{Intensity brightness maps}

Once a best spectral energy distribution (SED) model was selected for
each star, we retrieved from DUSTY the intensity brightness profile at
the LABoCa wavelength. Each 1D intensity profile was converted to the observed 2D 
pixel grid and was convolved
with a 2D PSF of the size of the experimental HPBW. We then compared the
synthetic brightness maps to the LABoCa maps.

For a better visualisation we used differential aperture photometry on
the data. What we expect to see when doing that for a source with a symmetric and homogeneous non detached CSE is a gaussian profile with a FWHM wider than the one obtained for a point source and representative of the extended emission. The gaussian profile would be disrupted if there is asymmetry and/or if there is any extra structure around the star such as a disc.

The background variation is the main source of photometric errors in
bolometric data. Any constant background would be subtracted with the
differential aperture photometry but its variation can still be seen
and would essentially affect the parts with a low signal (the wings of
the profile) and increase the photometric error estimation.

The comparison between the LABoCa maps and the synthetic maps can be
seen from Fig.~\ref{ocet} to Fig.~\ref{figLast}.
A good agreement is found between the synthetic map and the LABoCa map
for $o$\,Cet, CW\,Leo, AFGL\,3068.
For $\pi$\,Gru, the synthetic brightness map is close to the LABoCa
data up to 20$\arcsec$ but after that the intensity in the synthetic
map drops quickly to zero while we still have flux ($\sim$10mJy) in
the LABoCa map. This can be due to the asymmetry observed in the
LABoca map (see Fig.~\ref{PI_cont}) and which is not taken into
account in the modelling.
The synthetic brightness maps of IRC\,+10420, AFGL\,5379 and
IRAS\,16342$-$3814 predict more flux than the LABoCa maps up to 50$\arcsec$. 
At larger distance, the models are within the error bars of the LABoCa data.
For IRC\,+10420 and IRAS\,16342$-$3814 (Dijkstra et al. 2003) departure 
from spherical symmetry may be the cause.
WX\,Psc synthetic map shows an important discrepancy with the LABoCa map. 
The synthetic map predicts more flux than observed and a more
extended structure around the star, although the SED is well fitted.

The degree of agreement between the synthetic brightness maps and
the LABoCa maps does not seem to be related to the quality of the SED modelling.
Note that this discrepancy is not related to the assumed distance ($D$). Both the luminosity ($L$) 
and the angular scale ($R$) depend in the same way on the adopted distance ($L/D^2 = constant$,
while $L \sim R^2$, so $R/D = constant$).

In most cases, the models predict more dust than what is seen in the observed 
data. We tested decreasing the maximum value of the CSE outer boundary, but that 
did not make any difference up to a certain value and after that the quality of the SED
model became poor. This is not surprising, as the outer boundary set in DUSTY
is a maximum limit for the calculation of the dust and does not represent the final size of 
the CSE. The only risk would be to under estimate the limit, but that was not our case.

The main limitation in the modelling is the spherical symmetry assumption.  While it does not stop us from obtaining a decent fit to the SEDs even for the asymmetric sources it clearly shows its limitation when deriving synthetic intensity profiles. It is probably the reason we find a poor agreement between the synthetic brightness maps and the LABoCA intensity profiles. This clearly shows that we cannot rely only on SED modelling to properly constrain the physical parameters of CSEs. Combining both modelling of the SED and the intensity profile would be the ideal.

SABoCa data have been collected for WX\,Psc
and $o$\,Cet and still need to be analysed.
SABoCa is an APEX bolometer operating at 350$\mu$m with 
an expected sensitivity 5 times better than LABoCa.
These new data will have a better spatial resolution than 
the LABoCa data and should improve the results of this paper 
for these two stars.

\section{Conclusion}
\label{conclusion}

This paper presents the first LABoCa observations of evolved stars.
We show in this study that it is possible to resolve the extended
emission around AGB stars at 870 $\mu$m.

Extended emission has been found for CW\,Leo, $\pi$\,Gru, WX\,Psc and
$o$\,Cet, with an asymmetric structure around $\pi$\,Gru, WX\,Psc and $o$\,Cet.

Using SED modelling results, the mass-loss rate has been derived for 
all the stars with a reasonable agreement with previous literature data and
the total mass-loss has been estimated as well as the dust mass.

The SEDs and intensity profiles have been modelled with a 1D radiative transfer code. 
In most cases it is difficult to fit both constraints simultaneously.
For some stars, departures from spherical symmetry are known from other observations at 
different wavelengths and different spatial scales. The asymmetry may be the reason why 
it is not possible to fit both SEDs and intensity brightness maps in a consistent way 
given that we assumed spherical symmetry in the modelling. 
The current single dish observations still lack
the spatial resolution to really probe the dust structure at 870 $\mu$m.

\begin{acknowledgements}
D. Ladjal wants to thank L. Decin for a careful reading of the paper and for providing critical suggestions.
She also wants to thank F. Schuller for his precious help with the data reduction.
She is grateful to the the anonymous referee for the constructive comments which helped improve the paper.
This research has made use of the SIMBAD database, operated at CDS, Strasbourg, France.
\end{acknowledgements}

\clearpage
\newpage
\normalsize
\begin{appendix}

\section{LABoCa maps}

\begin{figure}
\vspace{0cm}
\hspace{0cm}
\resizebox{8cm}{!}{ \includegraphics{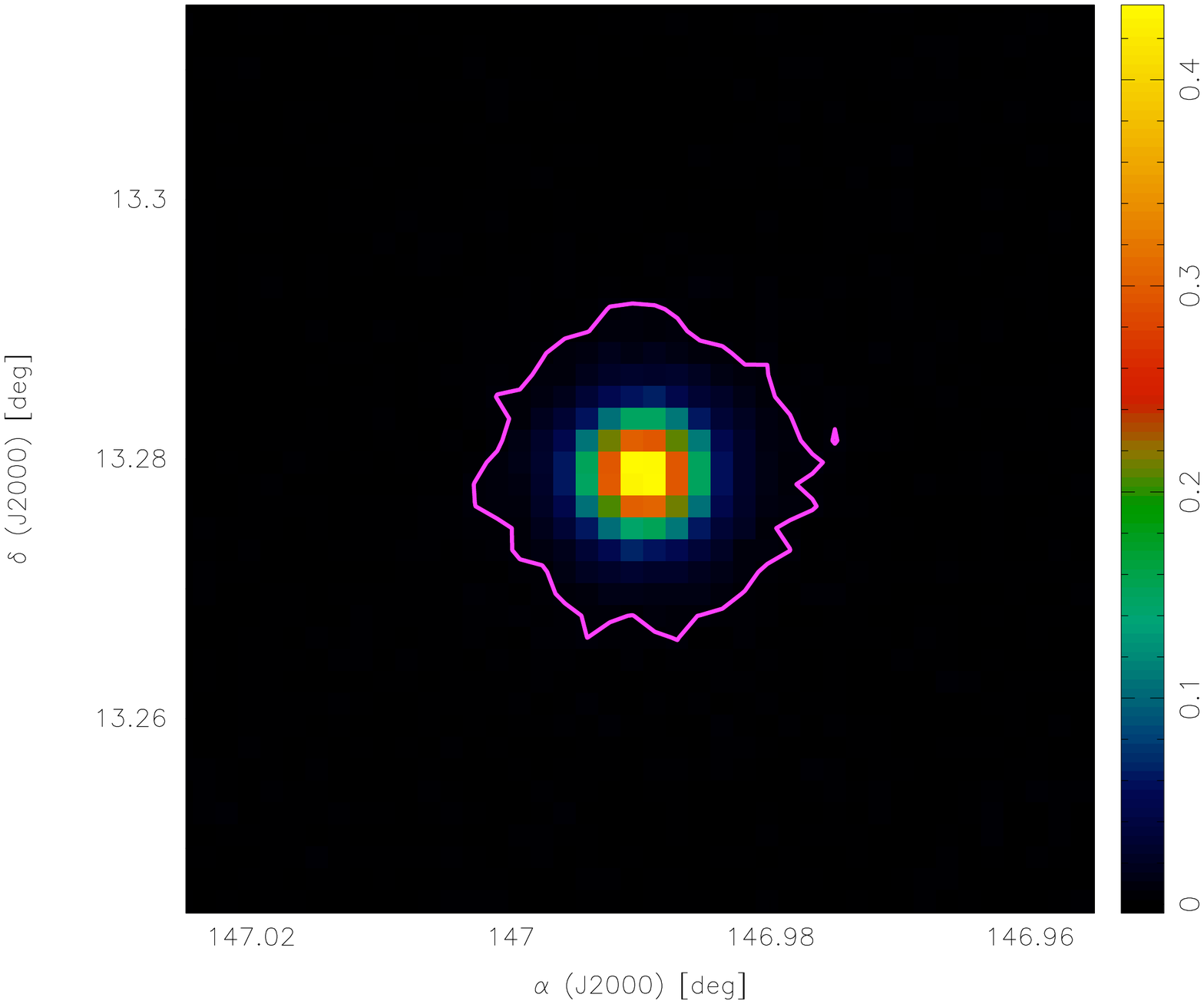}}
\caption{LABoCa brightness map for CW\,Leo in Jy (250\arcsec$\times$250\arcsec). The contour is drawn at a 3$\sigma$ level above the noise (7 MJy/sr).}
\label{CW_cont}

\vspace{0cm}
\hspace{0cm}
\resizebox{8cm}{!}{ \includegraphics{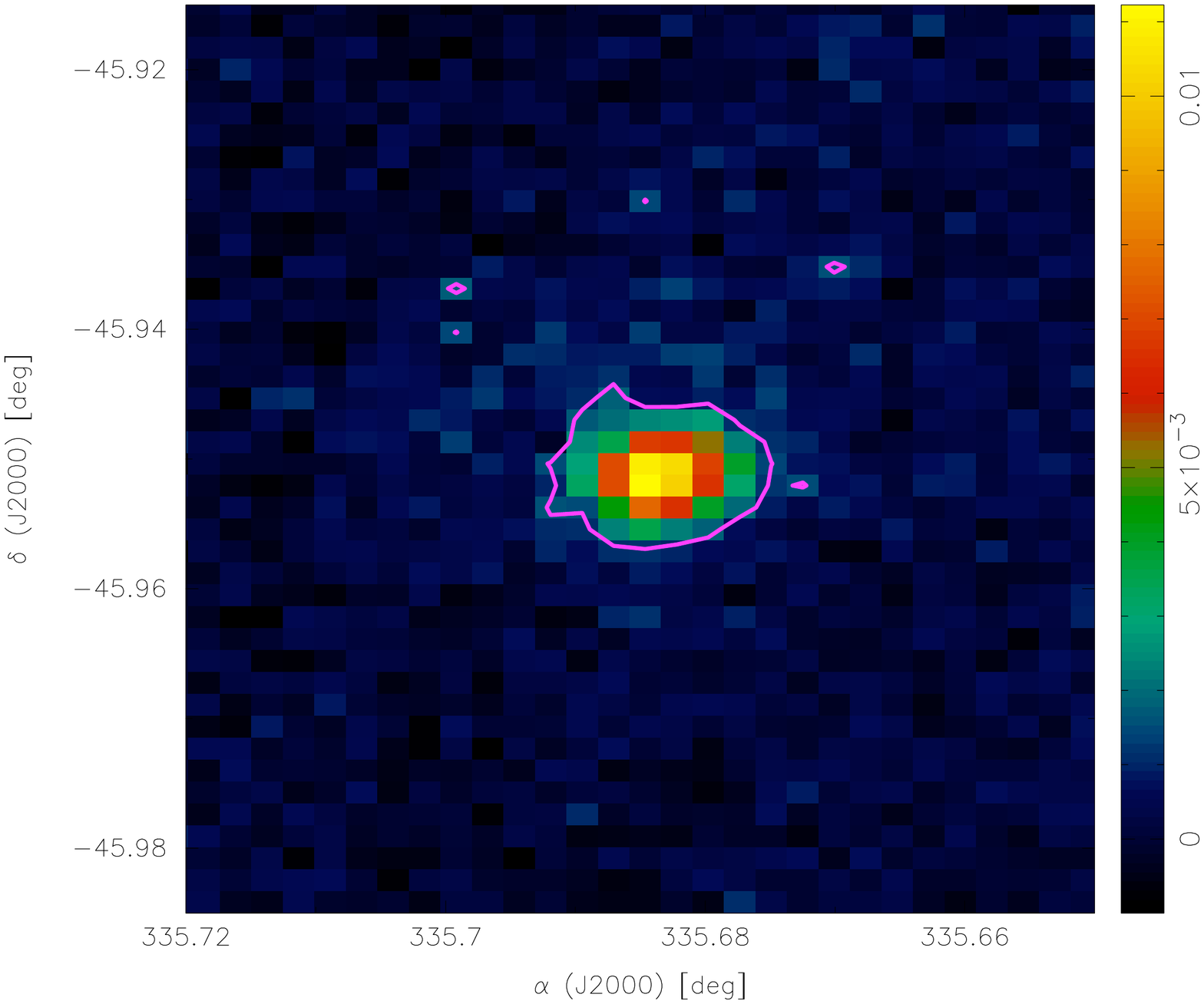}}
\caption{LABoCa brightness map for $\pi$\,Gru in Jy (250\arcsec$\times$250\arcsec). The contour is drawn at a 3$\sigma$ level above the noise (1.3 MJy/sr).}
\label{PI_cont}
\end{figure}

\begin{figure}
\vspace{0cm}
\hspace{0cm}
\resizebox{8cm}{!}{ \includegraphics{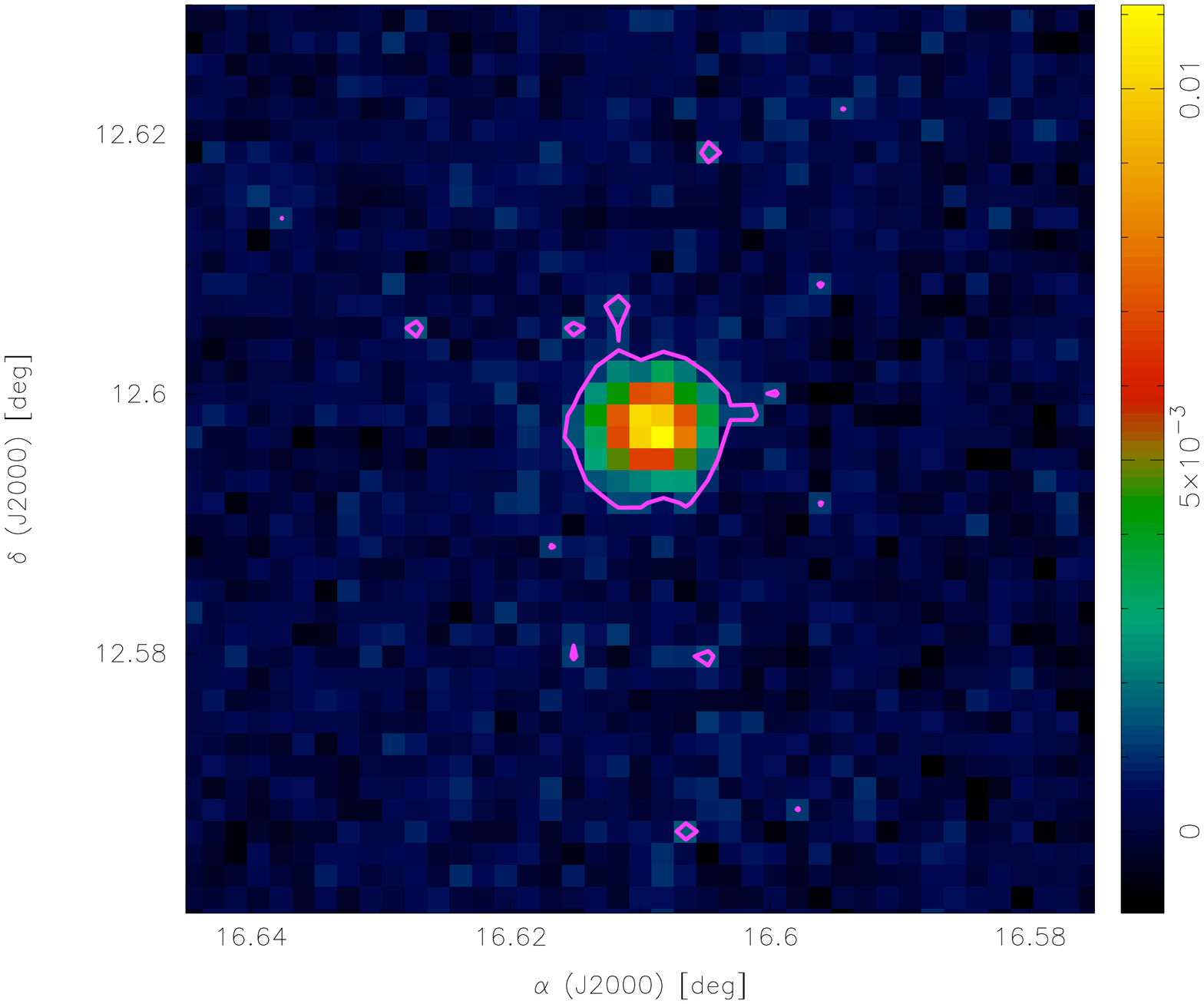}}
\caption{LABoCa brightness map for WX\,Psc in Jy (250\arcsec$\times$250\arcsec). The contour is drawn at a 3$\sigma$ level above the noise (1.3 MJy/sr).}
\label{WX_cont}

\vspace{0cm}
\hspace{0cm}
\resizebox{8cm}{!}{ \includegraphics{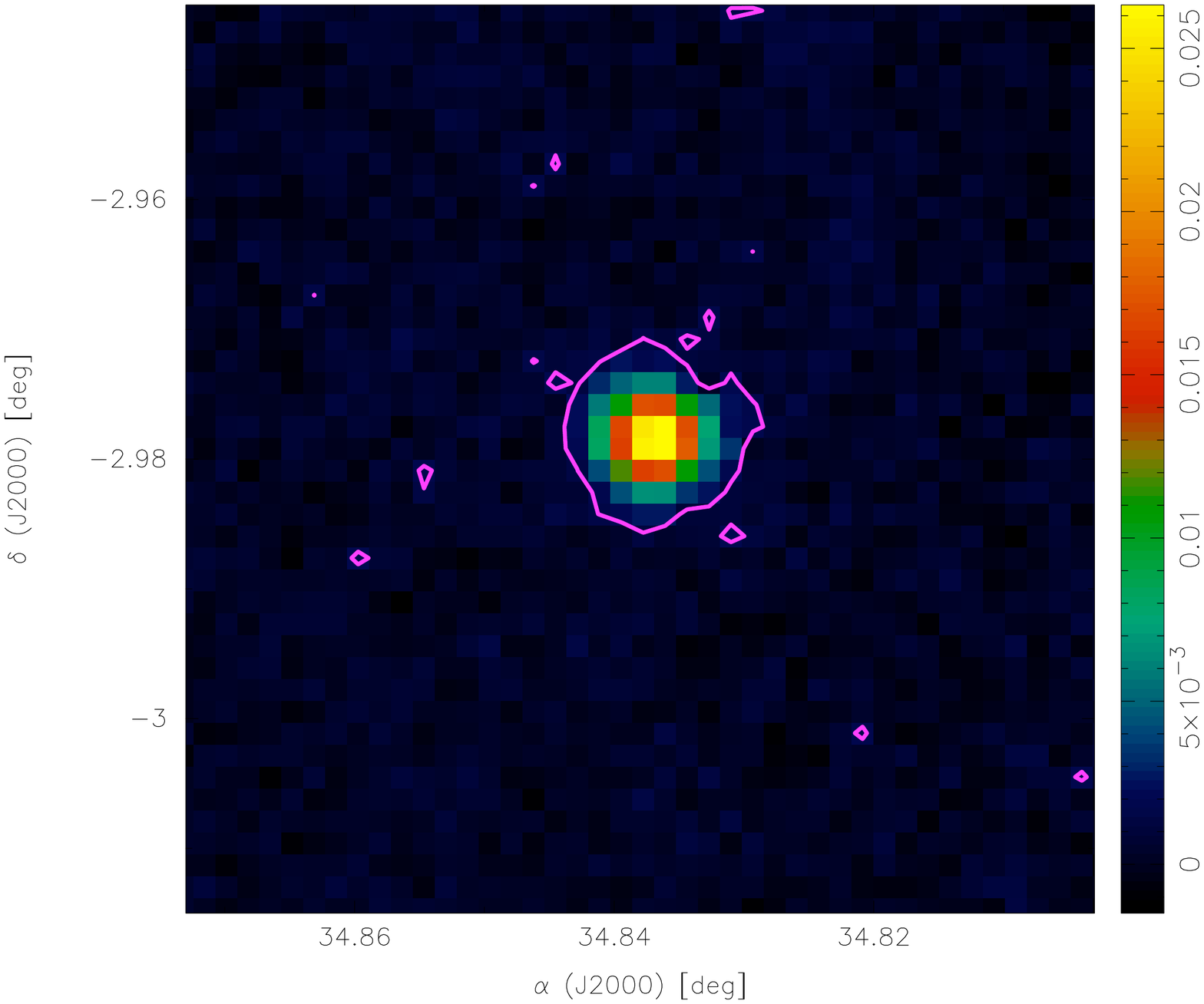}}
\caption{LABoCa brightness map for $o$\,Cet in Jy (250\arcsec$\times$250\arcsec). The contour is drawn at a 3$\sigma$ level above the noise (1.8 MJy/sr).}
\label{OMI_cont}
\end{figure}

\onecolumn
\section{Data reduction and calibration}
\label{obs}

\subsection{Data reduction}

The data reduction can be summarised as follows : 

In a first step, the data are corrected for the atmospheric opacity using skydip scans. Thereafter, they are converted from the raw units to Jy flux units (see the next section concerning the flux calibration). After flat-fielding, correlated noise is removed by subtracting the median noise across the array. The data are then filtered based on a low- and high-frequency cut-off. The final map is weighted by the inverse of the variance and gridded to a 6\arcsec\ pixel size.

Removing the correlated noise and filtering low frequencies are two very sensitive steps when looking for extended emission, because it is not possible to distinguish between a uniform extended emission and correlated noise observed by the bolometers. As a result, any uniform extended emission would be filtered out when performing the noise removal step.
For this reason, we decided to use an iterative reduction technique developed by F. Schuller at the Max-Planck-Insitut f\"{u}r Radioastronomy (MPIfR) which is meant to save the extended emission if present. The technique was successfully used in a (sub)millimetre survey of the galactic plane (see Schuller et al. 2009).

To execute this technique, we first perform the data reduction following the reduction steps described above. At the end of this first reduction, we use the resultant map to flag and select all signal above a 3~$\sigma$ level. We then remove that part of the signal and we re-run the reduction procedure on the residual signal to enhance any faint emission left. At the end of this step, we put back the subtracted flux. This constitutes one iteration block. The iteration is repeated several times until a convergence in the signal-to-noise is obtained. 

\subsection{Calibration}
\label{calibration}

BoA provides a flux calibration procedure which convert the maps from the raw units (Volts/beam) to Jy/beam. This procedure uses a Volt to Jy conversion factor specific to the instrument and a correction factor based on the ratio between the observed flux for a primary calibrator (a planet) and its predicted flux at the LABoCa frequency and beam and for a specific observing time.
BoA uses the primary calibrators to calibrate a list of secondary calibrator stars. 
The calibration factor for a science star is obtained by interpolating between the two calibration factors of the two calibrators (primary or secondary) bracketing the science object in time. The calibration is done to the peak flux.

This calibration method is adapted to point sources but if used with extended sources, only a fraction of the total flux is enclosed within the beam which would lead to an overestimation of the flux.
Therefor, we decided to develop our own reduction technique using only primary calibrators and by using aperture photometry. We chose to only use Uranus, Mars, Neptune and Venus observations as calibrators.

Using aperture photometry, we computed the integrated flux difference between two successive apertures in a map. This way, we take out any constant noise from the background. A 1D Gaussian is then fitted to the obtained profile. 

The Astro software from the Gildas package provided us with the expected total flux for each planet at the LABoCa frequency and for the right observing date.  The calibration factor is then the ratio between the expected total flux and the detected signal within an aperture of 3 sigma radius (corrected by the 99.73\% factor).

\section{Grid of models}
\label{grid}

After studying the effect of the different input parameters on the synthetic SED, we first constructed a grid of over 47,000 models by varying the following physical parameters :\\

- The optical depth $\mathbf{\tau}_{\lambda}$ at 0.55$\mu$m : Studying the effect of varying the optical depth on the resultant SED, it appeared that a change of about 10\% has a significant effect on the shape of the model. This means that a better sampling is needed at low optical depths than at high values. For this reason, we varied $\mathbf{\tau}_{\lambda}$ from 0.01 to 664 following a power law $\mathbf{\tau}_{\lambda}=0.01\times1.4^{i}$ (were $i$ is the index) which gives a finer sampling at low values.\\

- The effective temperature $T_{\rm eff}$ was varied from 2400~K to 3500~K, which is a typical temperature range for AGB stars, with a step size of 200~K. For IRC$+$10420 which is an F star, effective temperatures of 6000~K, 7000~K and 8000~K were tested.\\

- The dust condensation temperature $T_{\rm c}$ which determines the inner dust radius was varied from 500~K to 1200~K with a step size of 100~K.\\

- The dust chemical composition :\\
\indent\indent\indent$\bullet$ For O-rich stars :\\
\indent\indent\indent\indent- Iron ranging from 0\% to 30\% with a step size of 10\%.\\		
\indent\indent\indent\indent- Alumina ranging from 0\% to 60\% with a step size of 20\%.\\		
\indent\indent\indent\indent- Amorphous silicates = 100\%$-$Iron$-$Alumina.\\		
\indent\indent\indent$\bullet$ For C-rich stars :\\
\indent\indent\indent\indent- SiC ranging from 0\% to 15\% with a step size of 5\%.\\		
\indent\indent\indent\indent- MgS ranging from 0\% to 30\% with a step size of 10\%.\\		
\indent\indent\indent\indent- Amorphous carbon = 100\%$-$SiC$-$MgS.\\
		
The dust grain size was set to 0.05$\mu$m, the maximum outer radius of the shell was set to be 10,000 times the inner shell radius and the density distribution was taken $\propto r^{-2}$.

\vspace{6mm}

We compared each star to the models using a least square minimisation routine and a first solution was derived for each star. From this result we constructed a second grid of models with a more refined sampling of the physical parameters : 100~K for $T_{\rm eff}$, 50~K for $T_{\rm c}$, 1\% for Alumina, SiC and MgS and 0.5\% for Iron, the optical depth was varied with steps of about 10\% around $\mathbf{\tau}_{\lambda}$ of the first solution. This time the density distribution slope was also varied in order to match the observations, the power density index range was taken from -2.7 to -1.7 with a step size of 0.1. The observations were again compared to this second grid of models and a final solution was derived for each source (see Table~\ref{SEDfit}).

\newpage
\section{Tables}

\begin{table*}{}
\centering
\caption{List of observations.}
\label{observations}
\begin{tabular}{llrrr}
\hline
IRAS name & Alternative ID & Observing date & Number of scans & Total integration time [s] \\ 
\hline\hline
01037$+$1219 & WX\,Psc & 28-07-2007 & 3 & 105 \\
                                           & & 01-08-2007 & 6 & 210 \\
                                           & & 02-08-2007 & 10 & 350 \\
                                           & & 07-08-2008 & 9 & 315 \\
                                           & & 12-08-2008 & 2 & 70 \\
                                           & & 14-08-2008 & 9 & 315 \\
                                           & & 15-08-2008 & 6 & 210 \\
                                           \hline
                                           & & & \textbf{45} & \textbf{1575} \\
                                           \\

02168$-$0312 & $o$\,Cet & 28-07-2007 & 7 & 245 \\
                                          & & 02-06-2008 & 9 & 315 \\
                                          & & 03-06-2008 & 16 & 560 \\
                                          & & 04-06-2008 & 3 & 105 \\
                                          & & 15-08-2008 & 4 & 140 \\
                                          \hline
                                          & & & \textbf{39} & \textbf{1365} \\
                                          \\

07209$-$2540 & VY\,CMa & 25-02-2008 & 4 & 125  \\
				       & & 26-02-2008 & 2 & 70 \\
				       & & 27-02-2008 & 1 & 35 \\
				       \hline
				       & & & \textbf{7} & \textbf{230} \\
				       \\

09452$+$1330 & CW\,Leo & 24-02-2008 & 2 & 70 \\
				       & & 25-02-2008 & 4 & 125 \\
				       & & 26-02-2008 & 9 & 300 \\
				       & & 27-02-2008 & 18 & 528 \\
				       & & 28-02-2008 & 5 & 145 \\
				       & & 29-02-2008 & 3 &105 \\
				       \hline
				       & & & \textbf{41} & \textbf{1270} \\
				       \\

16342$-$3814 & $-$ & 28-07-2007 & 8 & 280 \\
                                      & & 15-08-2008 & 6 & 210 \\
                                      & & 16-08-2008 & 16 &560 \\
                                      \hline
                                      & & & \textbf{30} & \textbf{1050} \\
                                      \\

17411$-$3154 & AFGL\,5379 & 28-07-2007 & 8 & 280 \\ 
                                                    & & 15-08-2008 & 9 & 315 \\
                                                    & & 16-08-2008 & 4 & 140 \\
                                                    \hline
                                                    & & & \textbf{21} & \textbf{735} \\
                                                    \\

19244$+$1115 & IRC\,$+$10420 & 28-07-2007 & 7 & 245 \\
						    & & 13-08-2008 & 4 & 140 \\
						    & & 14-08-2008 & 23 & 805 \\
					              \hline
						    & & & \textbf{34} & \textbf{1190} \\
						    \\

22196$-$4612 & $\pi$\,Gru & 28-07-2007 & 4 & 140 \\
					  & & 01-08-2007 & 5 & 175 \\
					  & & 10-08-2008 & 18 & 630 \\
					  & & 11-08-2008 & 10 & 350 \\
					  & & 15-08-2008 & 2 & 70 \\
					  \hline
					  & & & \textbf{39} & \textbf{1365} \\
					  \\

23166$+$1655 & AFGL\,3068 & 28-07-2007 & \textbf{8} & \textbf{280} \\ 
\hline\hline
\end{tabular}
\end{table*}

\begin{table*}{}
\scriptsize
\centering
\caption{Photometric data.}
\label{photodata}
\begin{tabular}{lrrc|lrrc|lrrc}
\hline
Star & $\lambda$ & $F_{\nu}$ & $reference$ & Star & $\lambda$ & $F_{\nu}$ & $reference$ & Star & $\lambda$ & $F_{\nu}$ & $reference$ \\
\hline
  & [$\mu m$] & [Jy] &  &   & [$\mu m$] & [Jy] &  &   & [$\mu m$] & [Jy] &  \\
\hline\hline
$o$\,Cet 	&	0.55	&	229.94	&	(	1	)	&	CW\,Leo	&	1.25	&	2.3	&	(	21	)	&	WX\,Psc	&	1.23	&	3.17	&	(	1	)	\\
	&	0.55	&	9.76	&	(	9	)	&		&	1.25	&	2.7	&	(	2	)	&		&	1.25	&	1.69	&	(	2	)	\\
	&	0.7	&	1104.27	&	(	10	)	&		&	1.65	&	74.9	&	(	2	)	&		&	1.65	&	14.25	&	(	2	)	\\
	&	0.7	&	1138.92	&	(	1	)	&		&	1.65	&	39.6	&	(	21	)	&		&	2.17	&	86.52	&	(	2	)	\\
	&	0.88	&	2131.97	&	(	1	)	&		&	2.17	&	469	&	(	2	)	&		&	2.2	&	113.8	&	(	1	)	\\
	&	0.88	&	5871.93	&	(	1	)	&		&	2.2	&	388.8	&	(	21	)	&		&	3.5	&	220	&	(	3	)	\\
	&	0.9	&	5027.64	&	(	10	)	&		&	2.2	&	321.9	&	(	3	)	&		&	3.5	&	330.67	&	(	1	)	\\
	&	1.04	&	7193.59	&	(	10	)	&		&	3.5	&	3046.1	&	(	3	)	&		&	4.9	&	365.1	&	(	3	)	\\
	&	1.25	&	3128.15	&	(	2	)	&		&	3.6	&	9072	&	(	21	)	&		&	12	&	1120	&	(	4	)	\\
	&	1.25	&	9698.38	&	(	10	)	&		&	4.9	&	9010.7	&	(	3	)	&		&	12	&	646.9	&	(	3	)	\\
	&	1.25	&	3845.33	&	(	1	)	&		&	10.2	&	39188.4	&	(	21	)	&		&	25	&	929.3	&	(	3	)	\\
	&	1.25	&	9748.4	&	(	1	)	&		&	12	&	47500	&	(	6	)	&		&	25	&	921	&	(	4	)	\\
	&	1.65	&	4364.15	&	(	2	)	&		&	12	&	31773.4	&	(	3	)	&		&	34.6	&	498.5	&	(	5	)	\\
	&	2.17	&	5118.35	&	(	2	)	&		&	25	&	23100	&	(	6	)	&		&	60	&	212	&	(	4	)	\\
	&	2.2	&	11275.24	&	(	1	)	&		&	25	&	28610.8	&	(	3	)	&		&	60	&	175.3	&	(	3	)	\\
	&	2.2	&	5446.6	&	(	1	)	&		&	60	&	5650	&	(	6	)	&		&	100	&	68.8	&	(	3	)	\\
	&	2.2	&	5927.1	&	(	3	)	&		&	60	&	5585	&	(	3	)	&		&	100	&	67.5	&	(	4	)	\\
	&	2.25	&	11149.2	&	(	10	)	&		&	100	&	2017.5	&	(	3	)	&		&	400	&	3	&	(	14	)	\\
	&	3.5	&	4334.9	&	(	3	)	&		&	100	&	922	&	(	6	)	&		&	1200	&	0.16	&	(	19	)	\\
	&	3.5	&	10023.4	&	(	10	)	&		&	140	&	1051.5	&	(	3	)	&		&	1200	&	0.33	&	(	19	)	\\
	&	3.5	&	4606.73	&	(	1	)	&		&	377	&	35.2	&	(	22	)	&		&	1300	&	0.08	&	(	13	)	\\
																				\cline{9-12}									
	&	4.9	&	3220.5	&	(	3	)	&		&	400	&	32	&	(	14	)	&		&		&		&				\\
	&	5	&	5425.61	&	(	1	)	&		&	450	&	18.45	&	(	12	)	&	$\pi$\,Gru	&	0.55	&	8.5	&	(	1	)	\\
	&	12	&	2415.6	&	(	3	)	&		&	450	&	29	&	(	14	)	&		&	0.55	&	10.22	&	(	9	)	\\
	&	12	&	4880	&	(	6	)	&		&	450	&	33.2	&	(	16	)	&		&	0.7	&	157.21	&	(	1	)	\\
	&	25	&	2322.8	&	(	3	)	&		&	800	&	6.66	&	(	16	)	&		&	0.88	&	1602.43	&	(	1	)	\\
	&	25	&	2260	&	(	6	)	&		&	811	&	9.8	&	(	22	)	&		&	1.25	&	3079.56	&	(	2	)	\\
	&	60	&	301	&	(	6	)	&		&	850	&	9.45	&	(	12	)	&		&	1.25	&	2943.97	&	(	1	)	\\
	&	60	&	263.2	&	(	3	)	&		&	900	&	9	&	(	14	)	&		&	1.65	&	5795.62	&	(	2	)	\\
	&	100	&	108.9	&	(	3	)	&		&	1000	&	4	&	(	17	)	&		&	2.17	&	5812.06	&	(	2	)	\\
	&	100	&	88.4	&	(	6	)	&		&	1100	&	1.85	&	(	16	)	&		&	2.2	&	4657.21	&	(	1	)	\\
	&	450	&	1.09	&	(	12	)	&		&	1100	&	3.5	&	(	16	)	&		&	2.2	&	4552.7	&	(	3	)	\\
	&	850	&	0.35	&	(	12	)	&		&	1136	&	3.5	&	(	22	)	&		&	3.5	&	2952.2	&	(	3	)	\\
	&	1200	&	0.13	&	(	11	)	&		&	1200	&	1.58	&	(	11	)	&		&	4.9	&	1323.7	&	(	3	)	\\
	&	1200	&	0.18	&	(	19	)	&		&	1200	&	3.8	&	(	19	)	&		&	12	&	908	&	(	6	)	\\
	&	1300	&	0.12	&	(	13	)	&		&	1300	&	1.52	&	(	18	)	&		&	12	&	729.6	&	(	3	)	\\
\cline{1-4}										\cline{5-8}																			
	&		&		&				&		&		&		&				&		&	25	&	437	&	(	6	)	\\
	

VY\,CMa	&	0.55	&	2.1	&	(	9	)	&	IRAS\,16342$-$3814	&	1.25	&	0.04	&	(	2	)	&		&	25	&	506.2	&	(	3	)	\\
	&	0.7	&	27.74	&	(	10	)	&		&	1.65	&	0.06	&	(	2	)	&		&	60	&	77.3	&	(	6	)	\\
	&	1.25	&	118.6	&	(	2	)	&		&	2.17	&	0.1	&	(	2	)	&		&	60	&	86	&	(	3	)	\\
	&	1.65	&	239.8	&	(	2	)	&		&	12	&	16.2	&	(	6	)	&		&	100	&	23.3	&	(	6	)	\\
	&	2.17	&	510	&	(	2	)	&		&	25	&	200	&	(	6	)	&		&	100	&	40.9	&	(	3	)	\\
																				\cline{9-12}									
	&	4.9	&	3042.78	&	(	10	)	&		&	34.6	&	365.6	&	(	5	)	&		&		&		&				\\
	&	12	&	9920	&	(	6	)	&		&	60	&	290	&	(	6	)	&	AFGL\,3068	&	2.17	&	0.04	&	(	2	)	\\
	&	12	&	9278.1	&	(	3	)	&		&	100	&	139	&	(	6	)	&		&	12	&	690	&	(	4	)	\\
										\cline{5-8}																			
	&	25	&	6650	&	(	6	)	&		&		&		&				&		&	25	&	736	&	(	4	)	\\
	&	25	&	12345.5	&	(	3	)	&	IRC\,$+$10420	&	1.23	&	10.94	&	(	8	)	&		&	60	&	244	&	(	4	)	\\
	&	34.6	&	4387.5	&	(	5	)	&		&	1.25	&	10.38	&	(	2	)	&		&	100	&	72.7	&	(	4	)	\\
	&	60	&	1450	&	(	6	)	&		&	1.65	&	15.58	&	(	2	)	&		&	400	&	3	&	(	14	)	\\
	&	60	&	2318.4	&	(	3	)	&		&	2.15	&	25.52	&	(	8	)	&		&	450	&	1.61	&	(	12	)	\\
	&	100	&	331	&	(	6	)	&		&	2.17	&	23.94	&	(	2	)	&		&	800	&	0.3	&	(	15	)	\\
	&	100	&	2084.5	&	(	3	)	&		&	3.5	&	104.1	&	(	3	)	&		&	850	&	0.38	&	(	12	)	\\
	&	400	&	10	&	(	14	)	&		&	4.35	&	152.1	&	(	7	)	&		&	1100	&	0.2	&	(	15	)	\\
	&	450	&	9.5	&	(	12	)	&		&	4.9	&	179.2	&	(	3	)	&		&	1200	&	0.43	&	(	19	)	\\
																				\cline{9-12}									
	&	800	&	2.81	&	(	16	)	&		&	8.28	&	148.7	&	(	7	)	&		&		&		&				\\
	&	780	&	1.99	&	(	20	)	&		&	12	&	1280.8	&	(	3	)	&	AFGL\,5379	&	1.25	&	0.01	&	(	2	)	\\
	&	780	&	2.66	&	(	20	)	&		&	12	&	1350	&	(	6	)	&		&	1.65	&	0.02	&	(	2	)	\\
	&	850	&	2.18	&	(	20	)	&		&	12.13	&	1292	&	(	7	)	&		&	2.17	&	0.04	&	(	2	)	\\
	&	850	&	1.77	&	(	12	)	&		&	14.65	&	1263	&	(	7	)	&		&	4.35	&	158.5	&	(	7	)	\\
	&	1100	&	0.79	&	(	20	)	&		&	21.34	&	2221	&	(	7	)	&		&	8.28	&	173.8	&	(	7	)	\\
	&	1100	&	0.92	&	(	20	)	&		&	25	&	2911.5	&	(	3	)	&		&	12	&	1260	&	(	4	)	\\
	&	1100	&	0.83	&	(	16	)	&		&	25	&	2310	&	(	6	)	&		&	12.13	&	1318	&	(	7	)	\\
	&	1200	&	0.34	&	(	11	)	&		&	60	&	718	&	(	6	)	&		&	14.65	&	1204	&	(	7	)	\\
	&	1200	&	0.69	&	(	19	)	&		&	60	&	566.9	&	(	3	)	&		&	21.34	&	1679	&	(	7	)	\\
	&	1250	&	0.79	&	(	20	)	&		&	100	&	186	&	(	6	)	&		&	25	&	2720	&	(	4	)	\\
	&	1250	&	0.58	&	(	20	)	&		&	450	&	0.95	&	(	12	)	&		&	60	&	1360	&	(	4	)	\\
	&	1300	&	0.38	&	(	13	)	&		&	850	&	0.31	&	(	12	)	&		&	100	&	406	&	(	4	)	\\
	&	1840	&	0.42	&	(	20	)	&		&	1200	&	0.5	&	(	19	)	&		&		&		&				\\
	&	1840	&	0.27	&	(	20	)	&		&	1300	&	0.1	&	(	13	)	&		&		&		&				\\
\hline\hline

\end{tabular}
\begin{tablenotes}[para]
References.
(1)	UBVRIJKLMNH Photoelectric Catalogue (Morel+ 1978);
(2)	2Mass All-Sky catalogue (Cutri+ 2003);
(3)	COBE DIRBE point source catalogue (Smith+ 2004);
(4)	IRAS faint source catalogue, version 2.0 (Moshir+ 1989);
(5)	The 35um absorption line towards 1612MHz masers (He+ 2005);
(6)	IRAS catalogue of point sources, version 2.0 (IPAC 1986);
(7)	MSX6C Infrared point source catalogue (Egan+ 2003);
(8)	DENIS photometry of bright southern stars (Kimeswenger+ 2004);
(9)	The Hipparcos and Tycho catalogues (ESA 1997);
(10)	Catalogue of late type stars with maser emission (Engels 1979);
(11)	Radio emission from stars at 250GHz (Altenhoff+ 1994);
(12)	Submillimeter Ð continuum SCUBA detections (Di Francesco+ 2008);
(13)	Walmsley et al. 1991;
(14)	Sopka et al. 1985;
(15)	Groenewegen et al. 1993;
(16)	Marshall et al. 1992;
(17)	Campbell et al. 1976;
(18)	Groenewegen et al. 1997;
(19)	Dehaes et al. 2007;
(20)	Knapp et al. 1993;
(21)	Le Bertre 1987;
(22)	Phillips et al. 1982.
\end{tablenotes}

\end{table*}

\onecolumn

\section{SED modelling}

\begin{figure}[htp]
\center
\begin{minipage}[c]{.48\linewidth}
\includegraphics[angle=90,width=8.6cm]{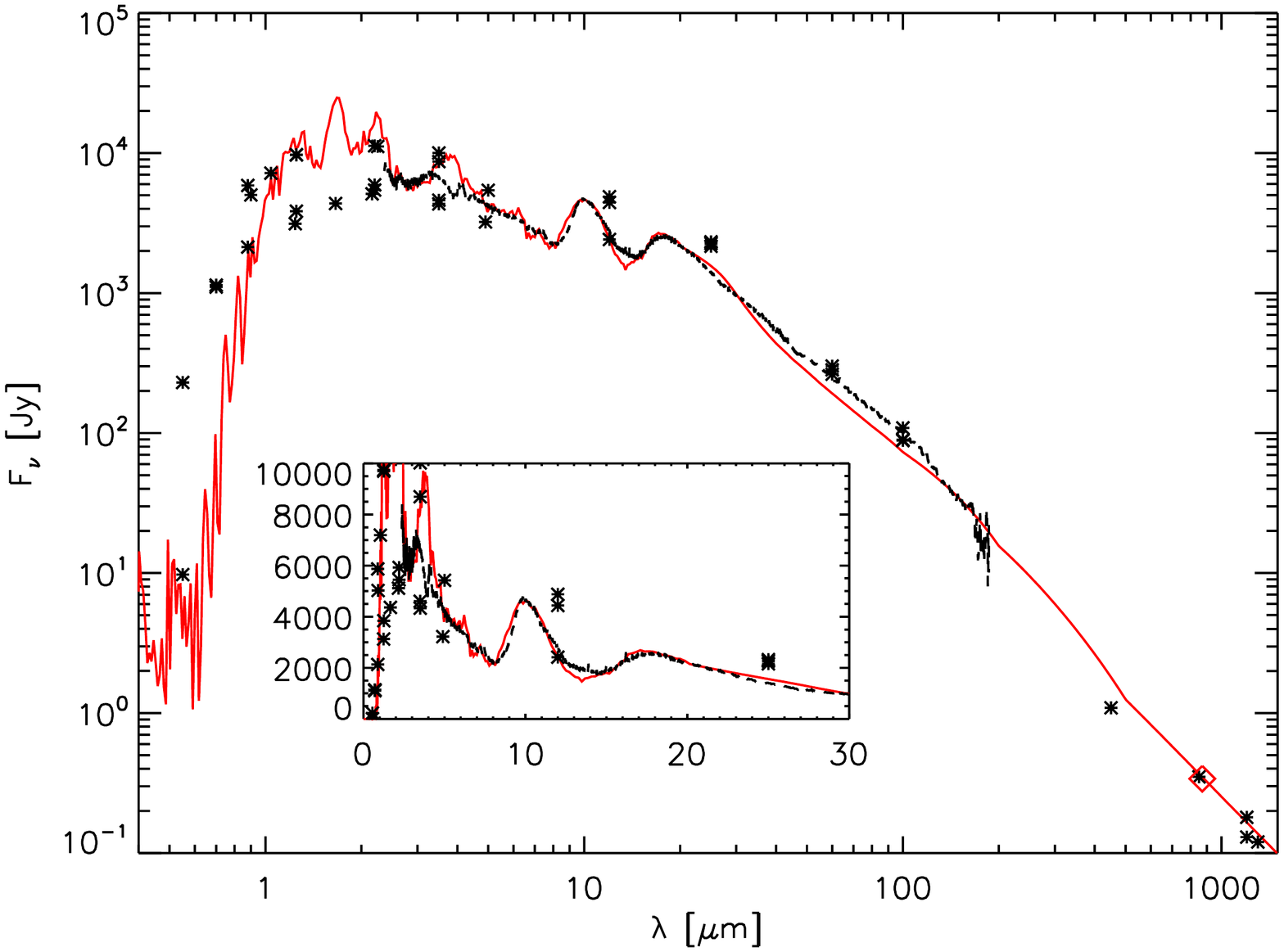}
\end{minipage}
\begin{minipage}[c]{.48\linewidth}
\includegraphics[angle=90,width=8.6cm]{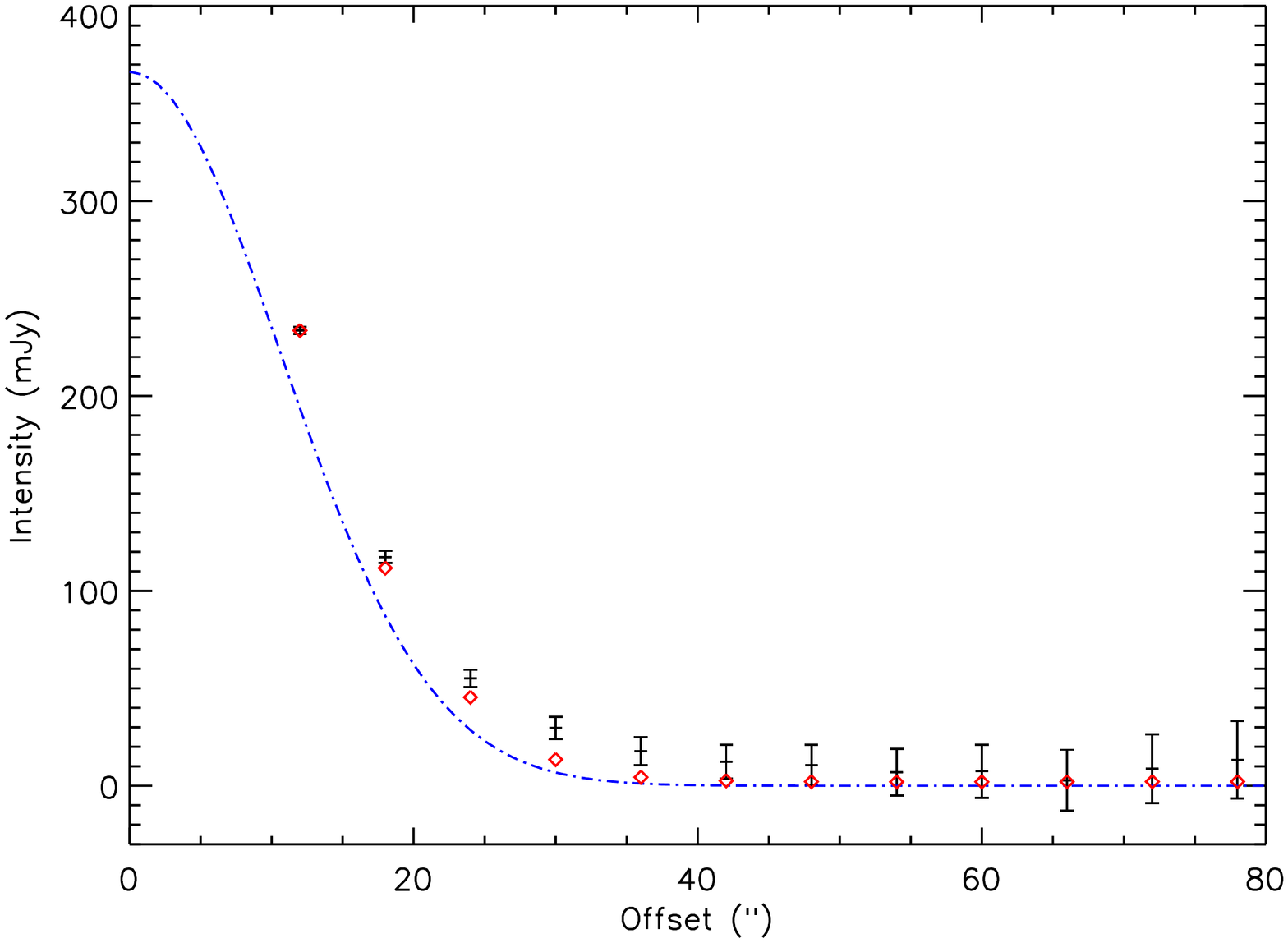}
\end{minipage}
\caption{
 Left panel: 
 best model (continuous red line) for ISO-SWS/LWS data of $o$\,Cet (dashed black line)
 with photometric data (black stars). The red diamond is the LABoCa
 measurement at 870 $\mu$m.
 The inset shows the spectral range between 0 to 50 $\mu$m.
 Right panel:
 Differential aperture photometry on the LABoCa map of $o$\,Cet
 (black symbols) compared to the synthetic brightness map derived
 from the best model (red symbols). 
 The dashed blue line is the HPBW.}
\label{ocet}

\begin{minipage}[c]{.48\linewidth}
\includegraphics[angle=90,width=8.6cm]{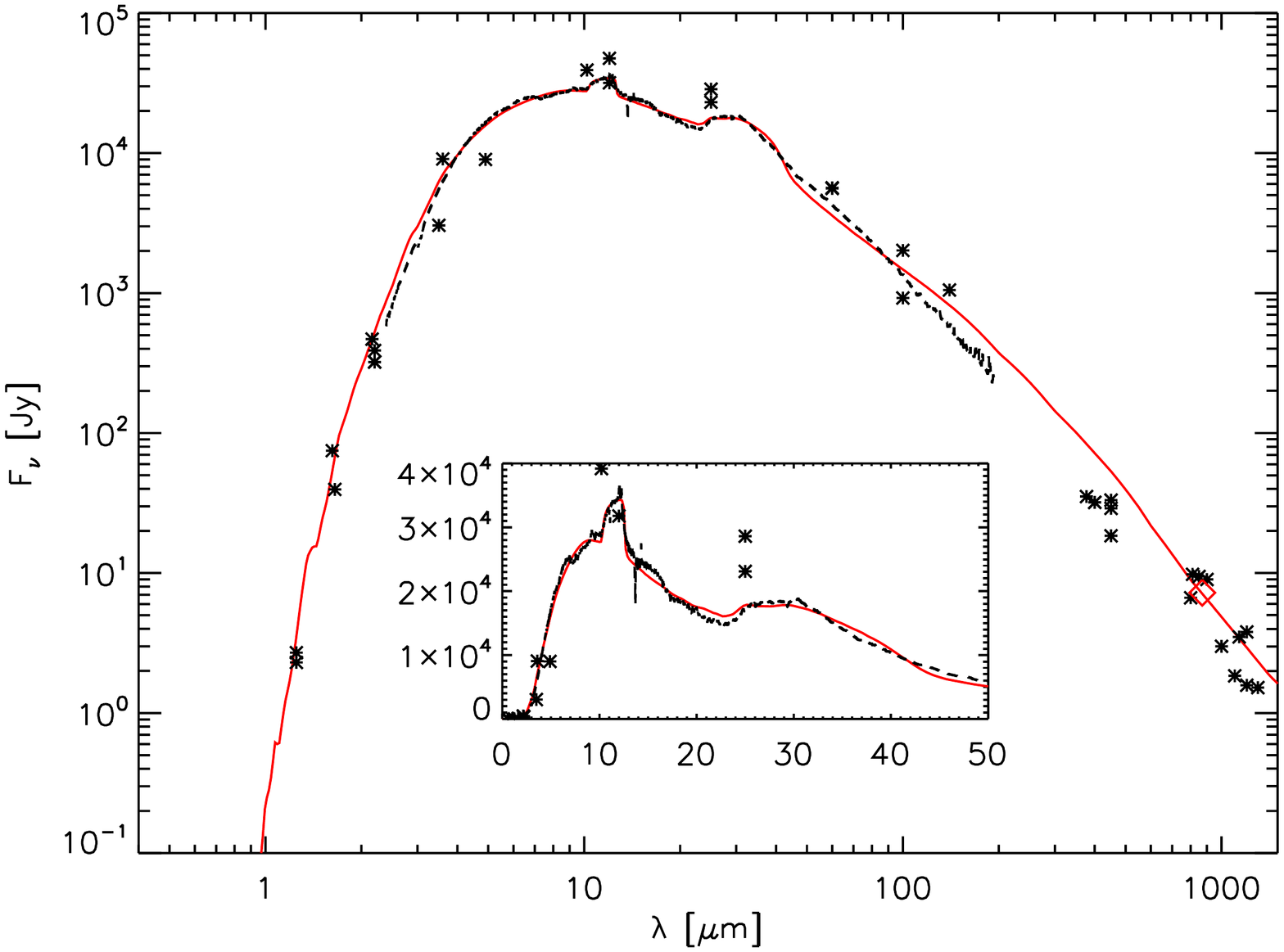}
\end{minipage}
\begin{minipage}[c]{.48\linewidth}
\includegraphics[angle=90,width=8.6cm]{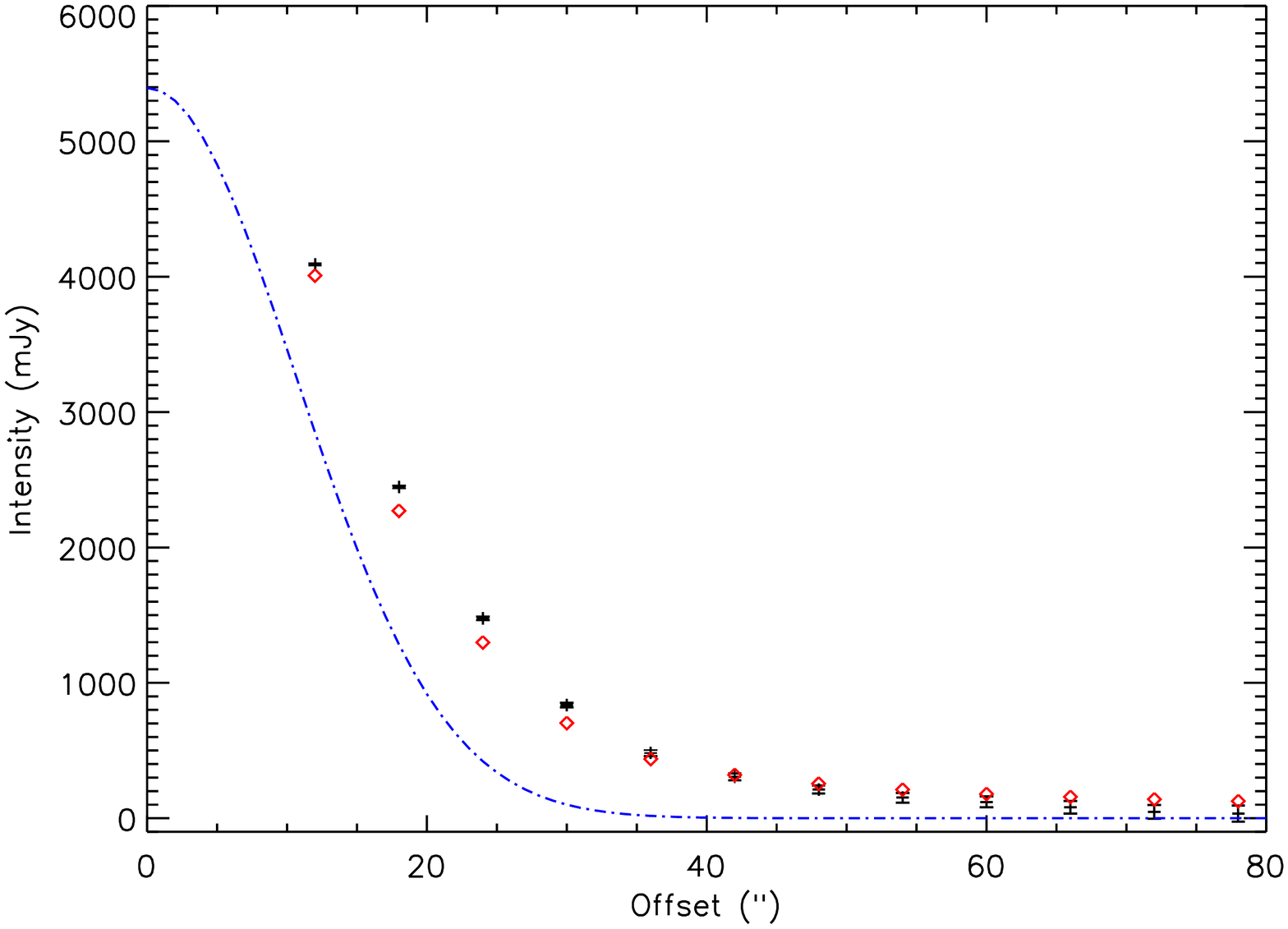}
\end{minipage}
\caption{As Figure~\ref{ocet} for CW Leo.
}

\begin{minipage}[c]{.48\linewidth}
\includegraphics[angle=90,width=8.6cm]{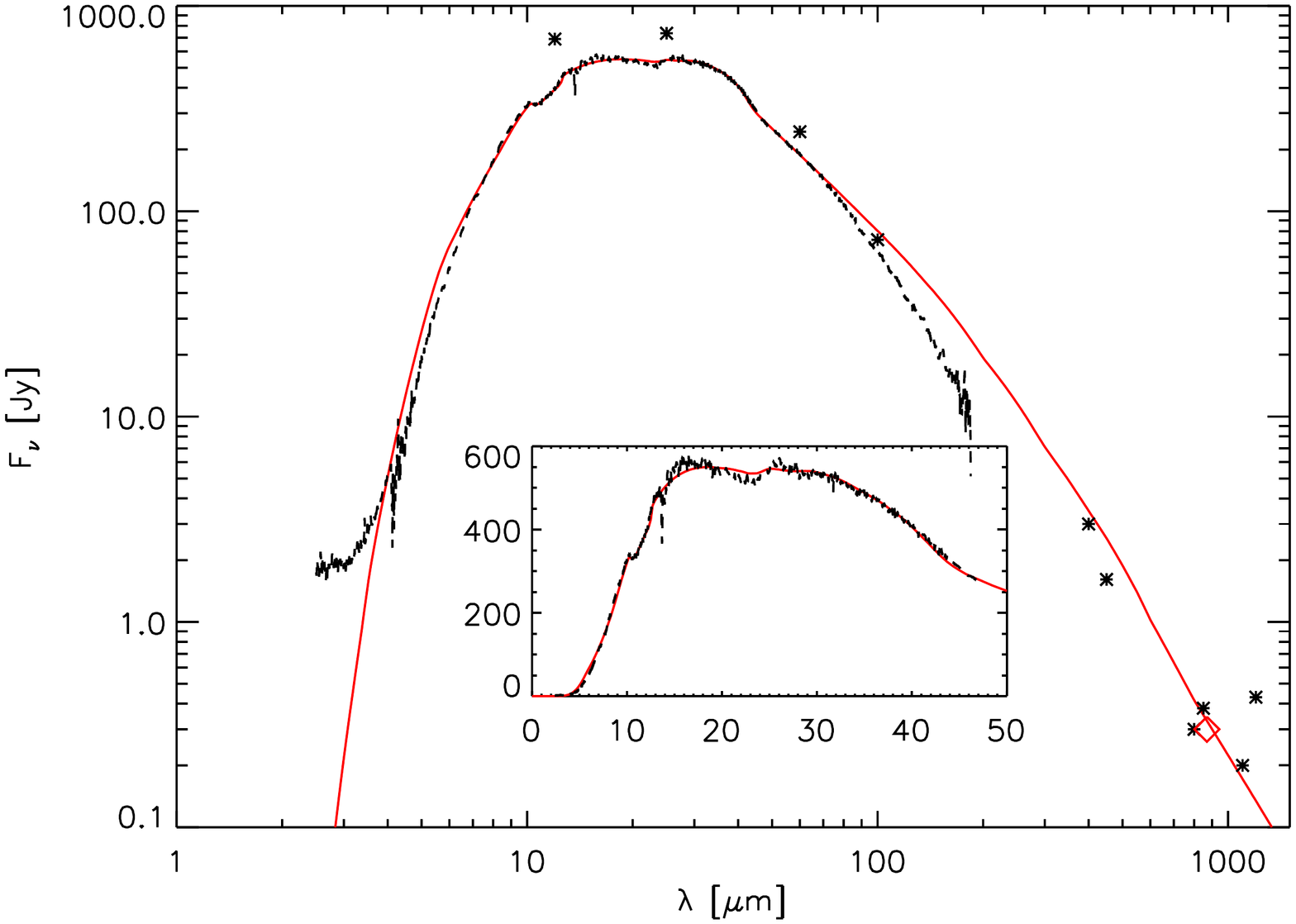}
\end{minipage}
\begin{minipage}[c]{.48\linewidth}
\includegraphics[angle=90,width=8.6cm]{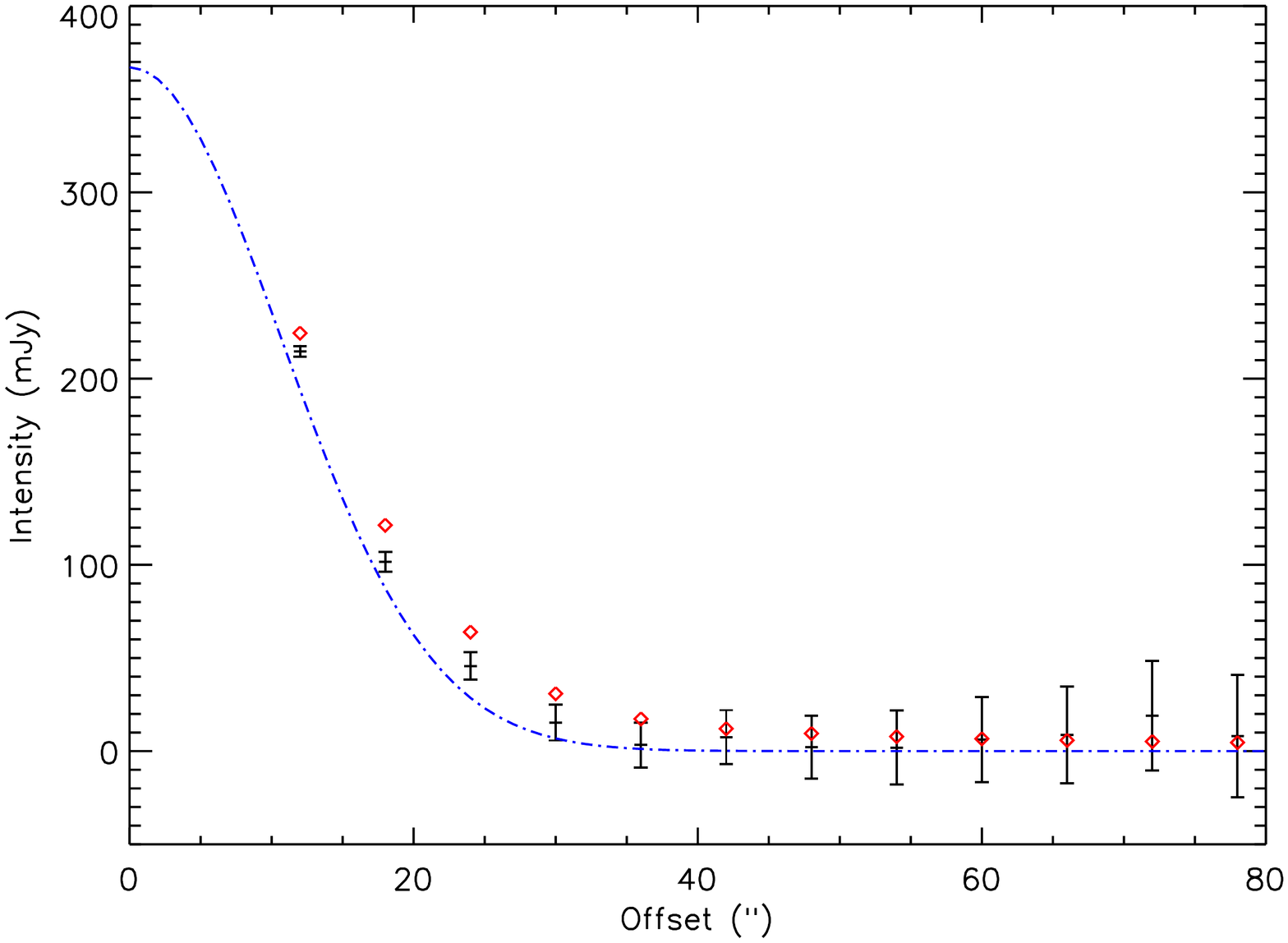}
\end{minipage}
\caption{As Figure~\ref{ocet} for  AFGL\,3068}
\label{figA}
\end{figure}


\begin{figure}[htp]
\center

\begin{minipage}[c]{.48\linewidth}
\includegraphics[angle=90,width=8.6cm]{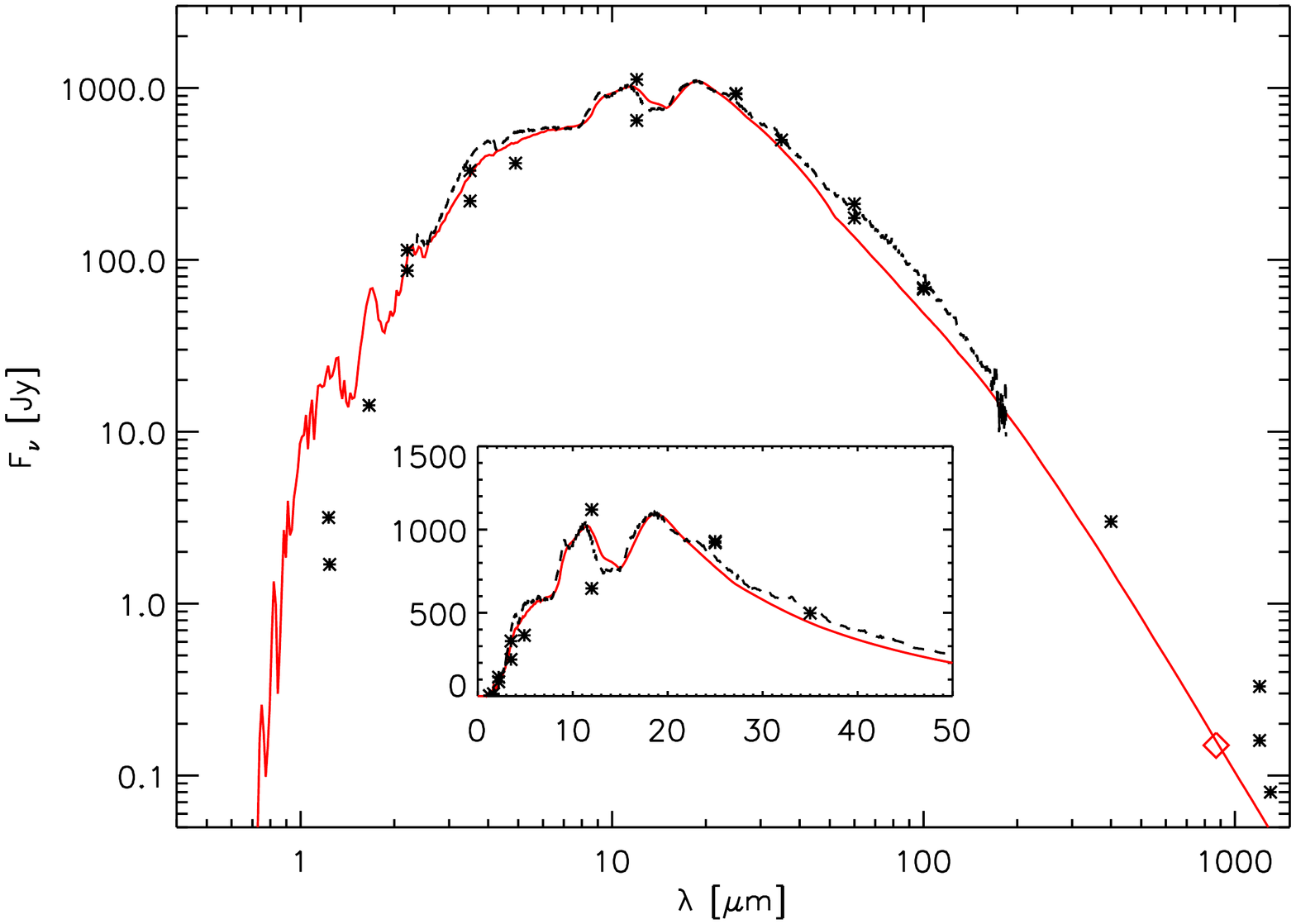}
\end{minipage}
\begin{minipage}[c]{.48\linewidth}
\includegraphics[angle=90,width=8.6cm]{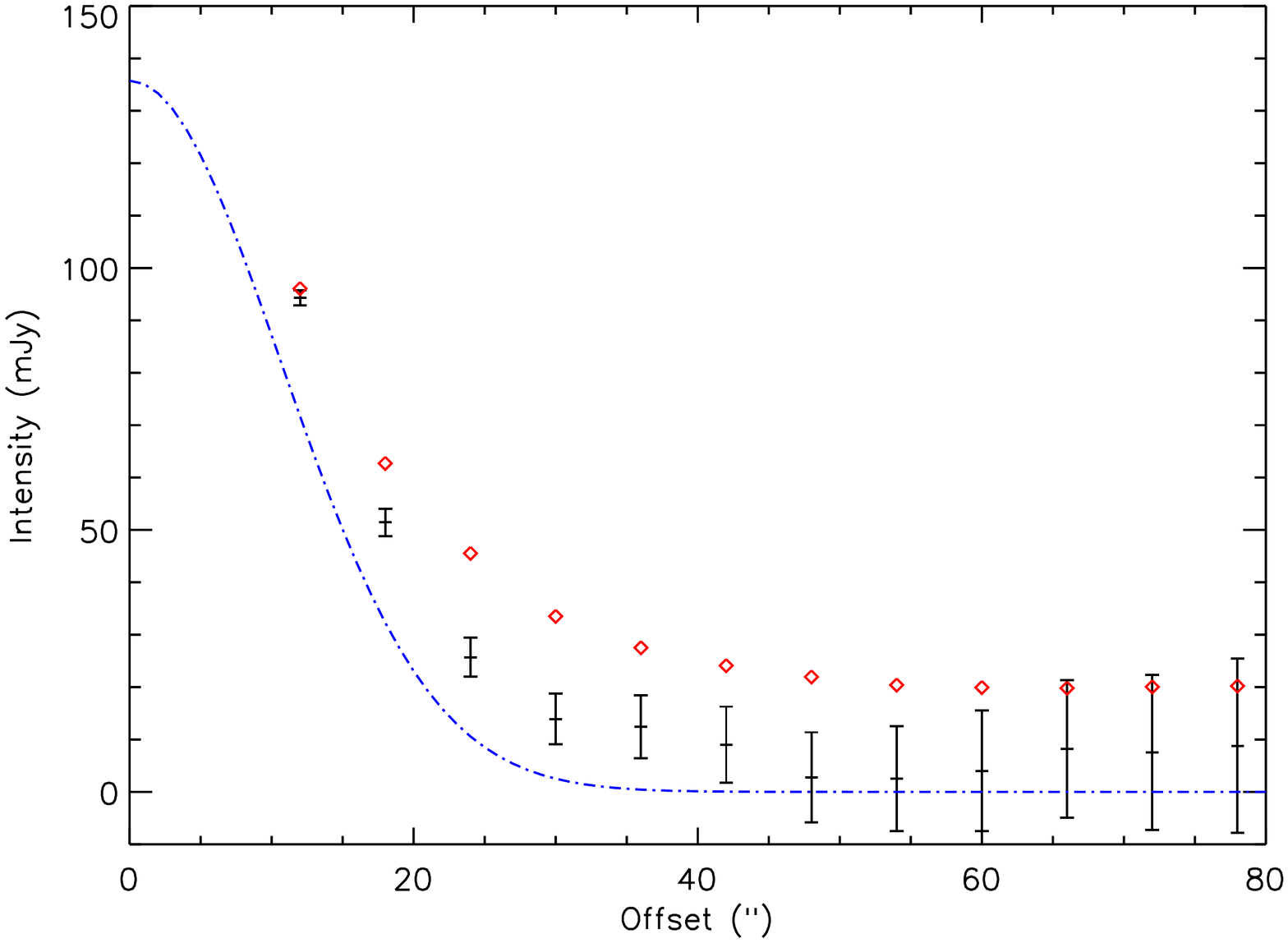}
\end{minipage}
\caption{As Figure~\ref{ocet} for WX\,Psc.}

\begin{minipage}[c]{.48\linewidth}
\includegraphics[angle=90,width=8.6cm]{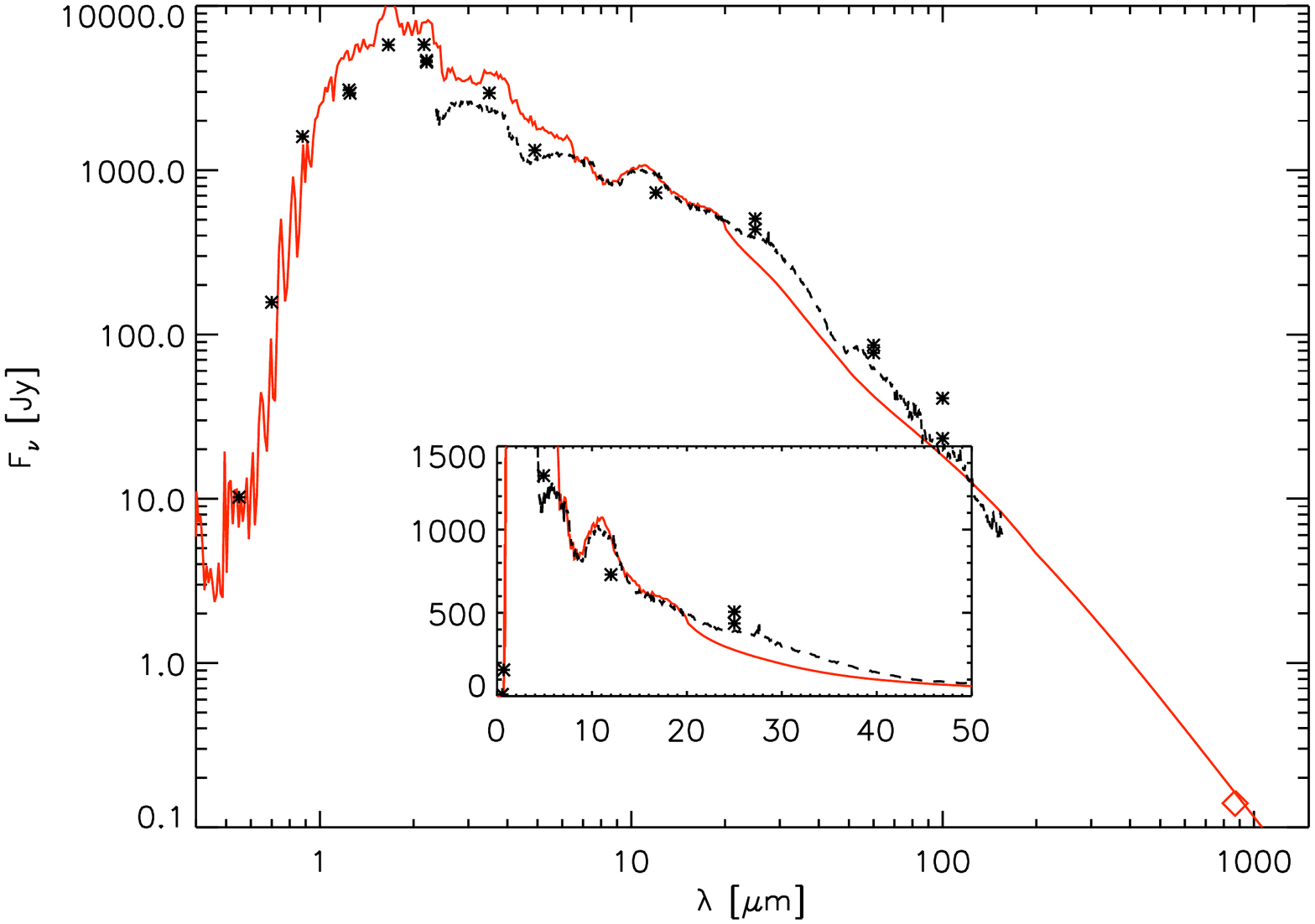}
\end{minipage}
\begin{minipage}[c]{.48\linewidth}
\includegraphics[angle=90,width=8.6cm]{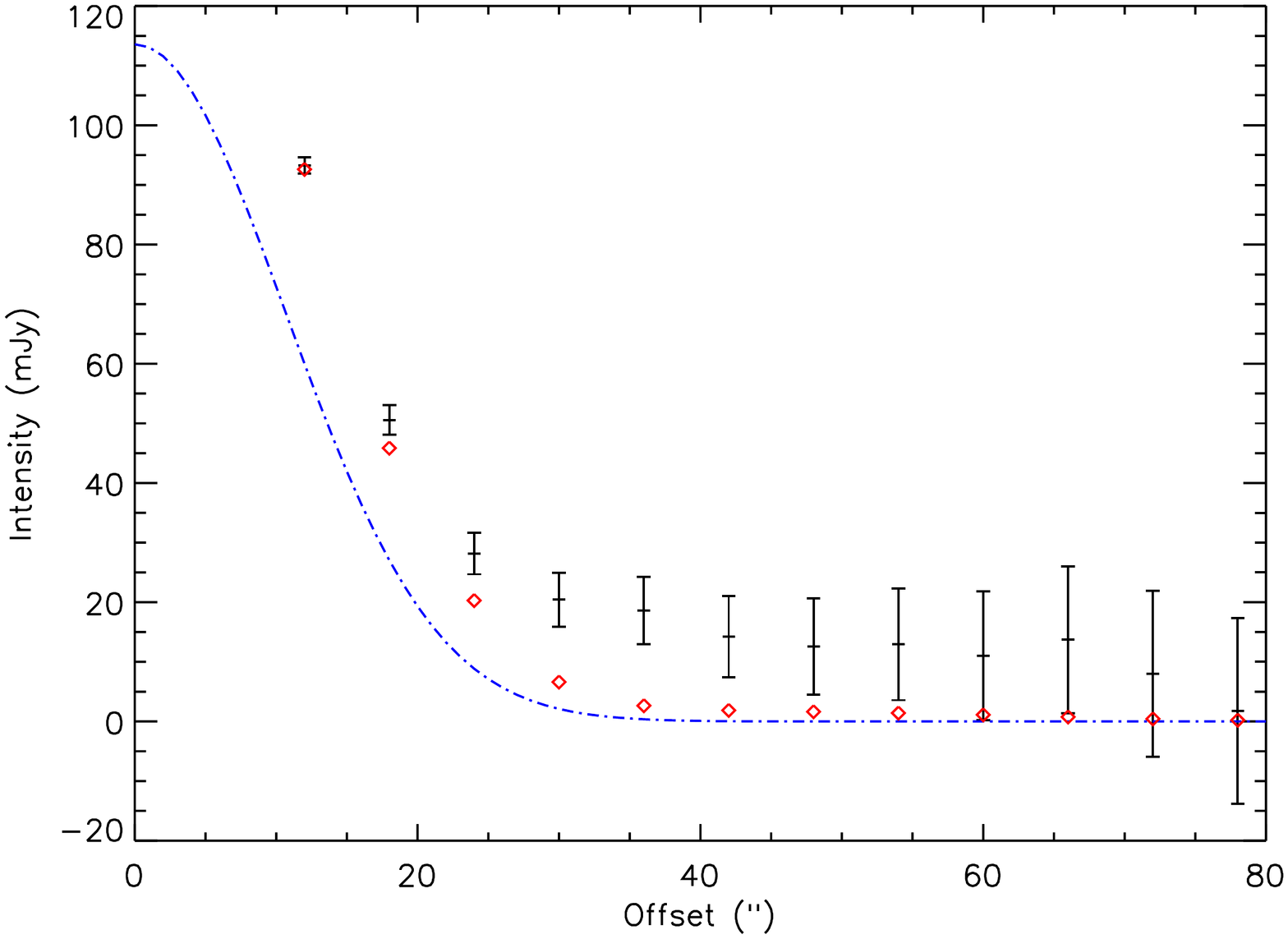}
\end{minipage}
\caption{As Figure~\ref{ocet} for $\pi$\,Gru.}

\begin{minipage}[c]{.48\linewidth}
\includegraphics[angle=90,width=8.6cm]{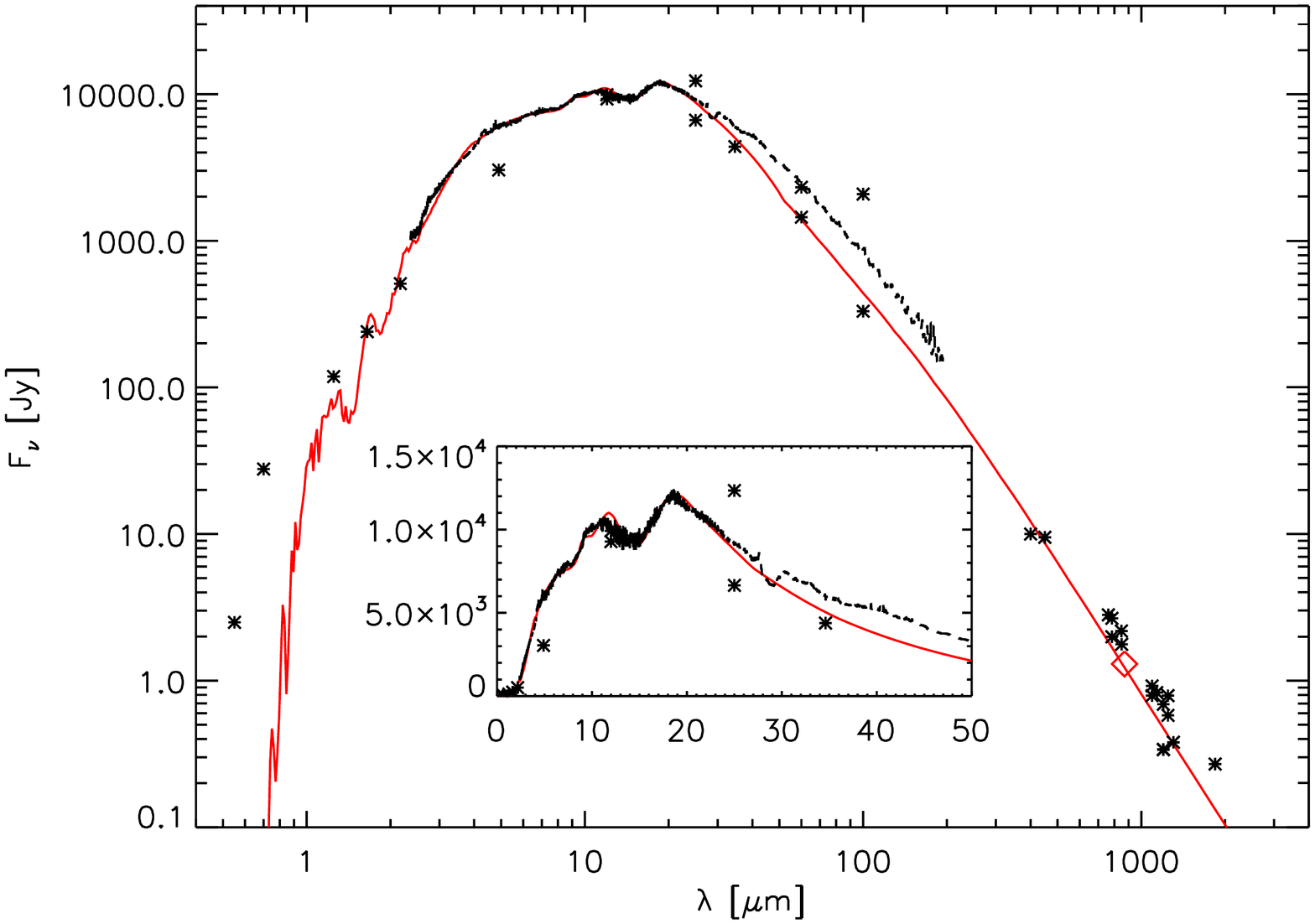}
\end{minipage}
\begin{minipage}[c]{.48\linewidth}
\includegraphics[angle=90,width=8.6cm]{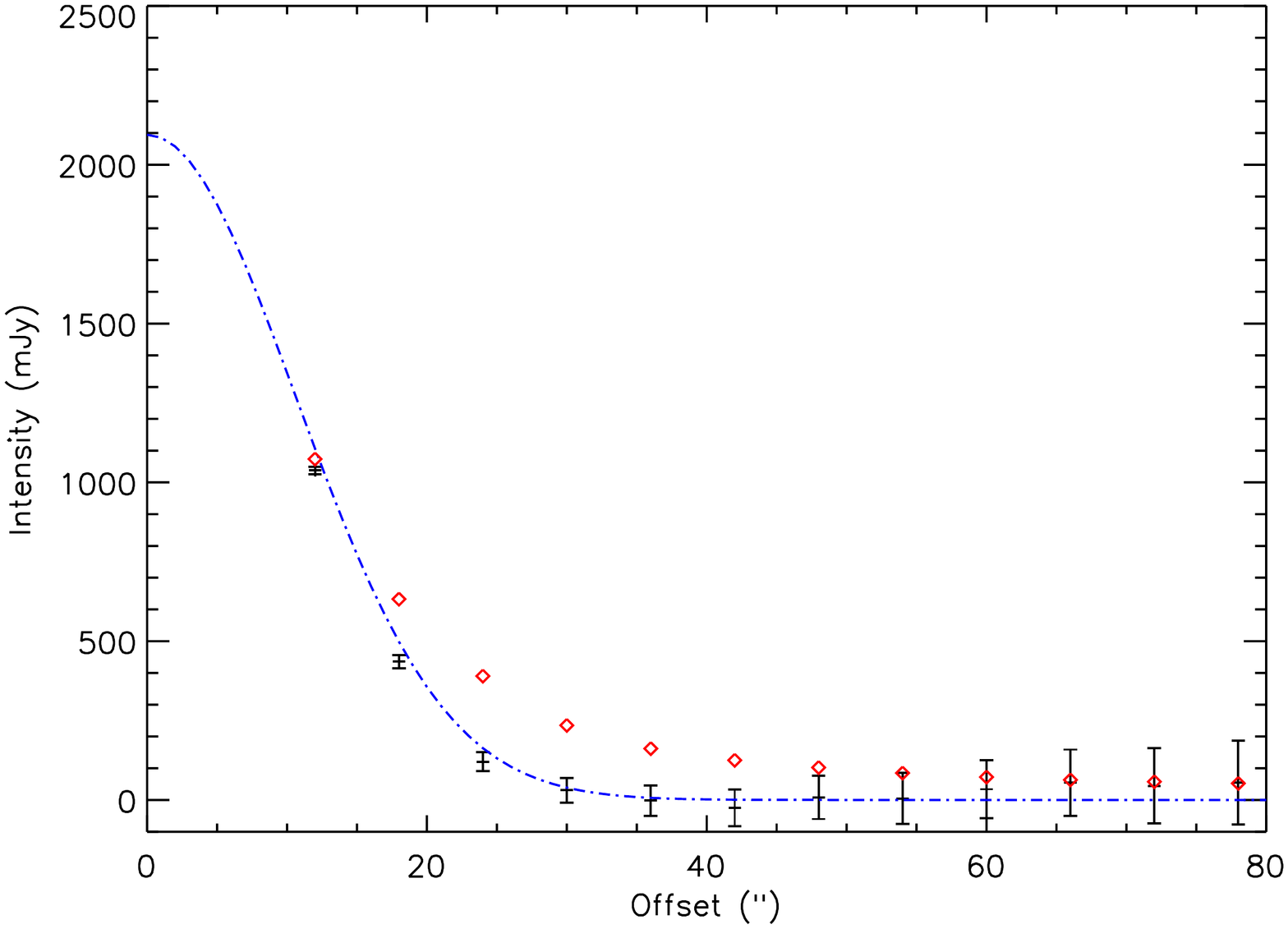}
\end{minipage}
\caption{As Figure~\ref{ocet} for VY\,CMa.}

\label{figB}
\end{figure}

\begin{figure}[htp]
\center

\begin{minipage}[c]{.48\linewidth}
\includegraphics[angle=90,width=8.6cm]{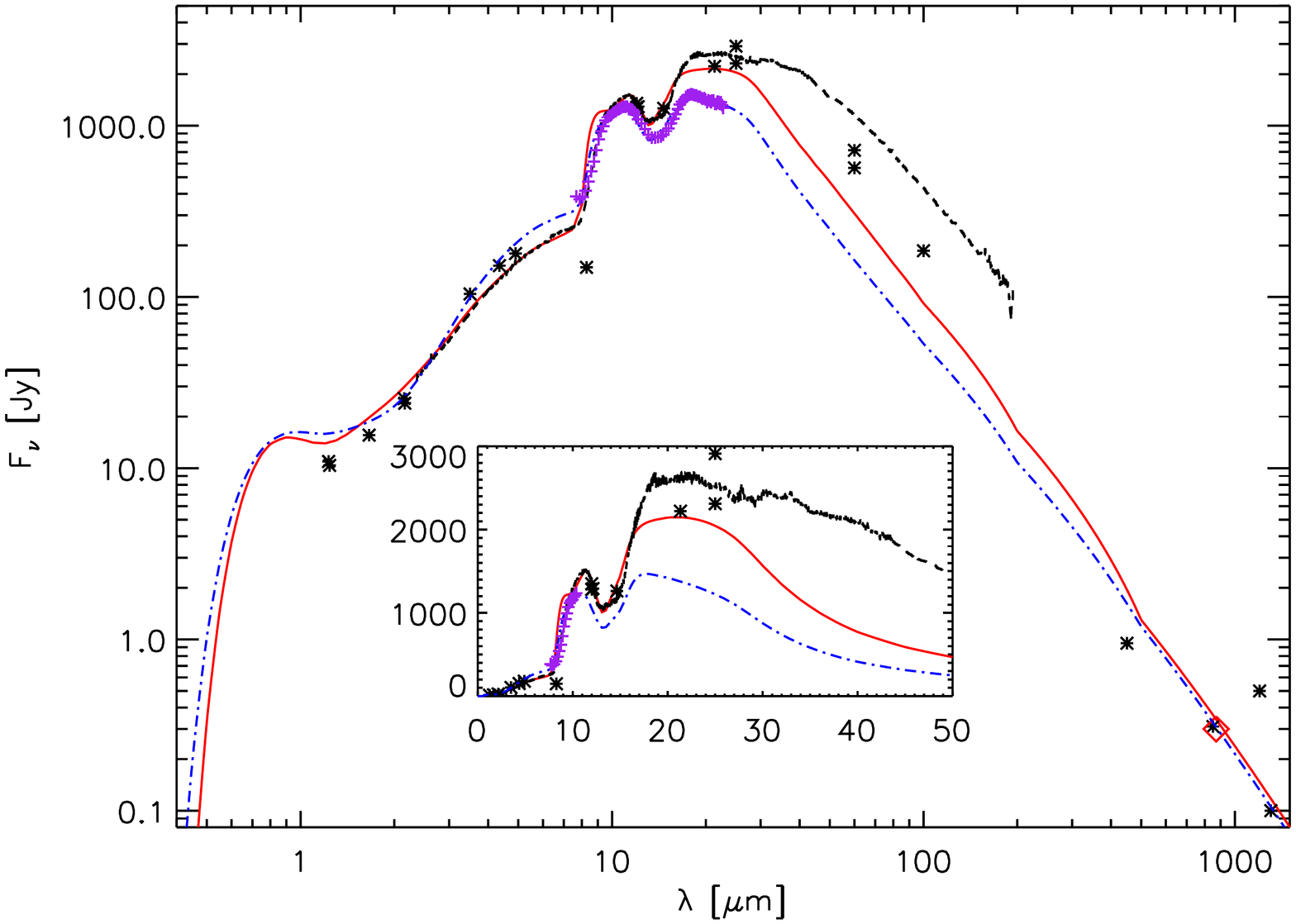}
\end{minipage}
\begin{minipage}[c]{.48\linewidth}
\includegraphics[angle=90,width=8.6cm]{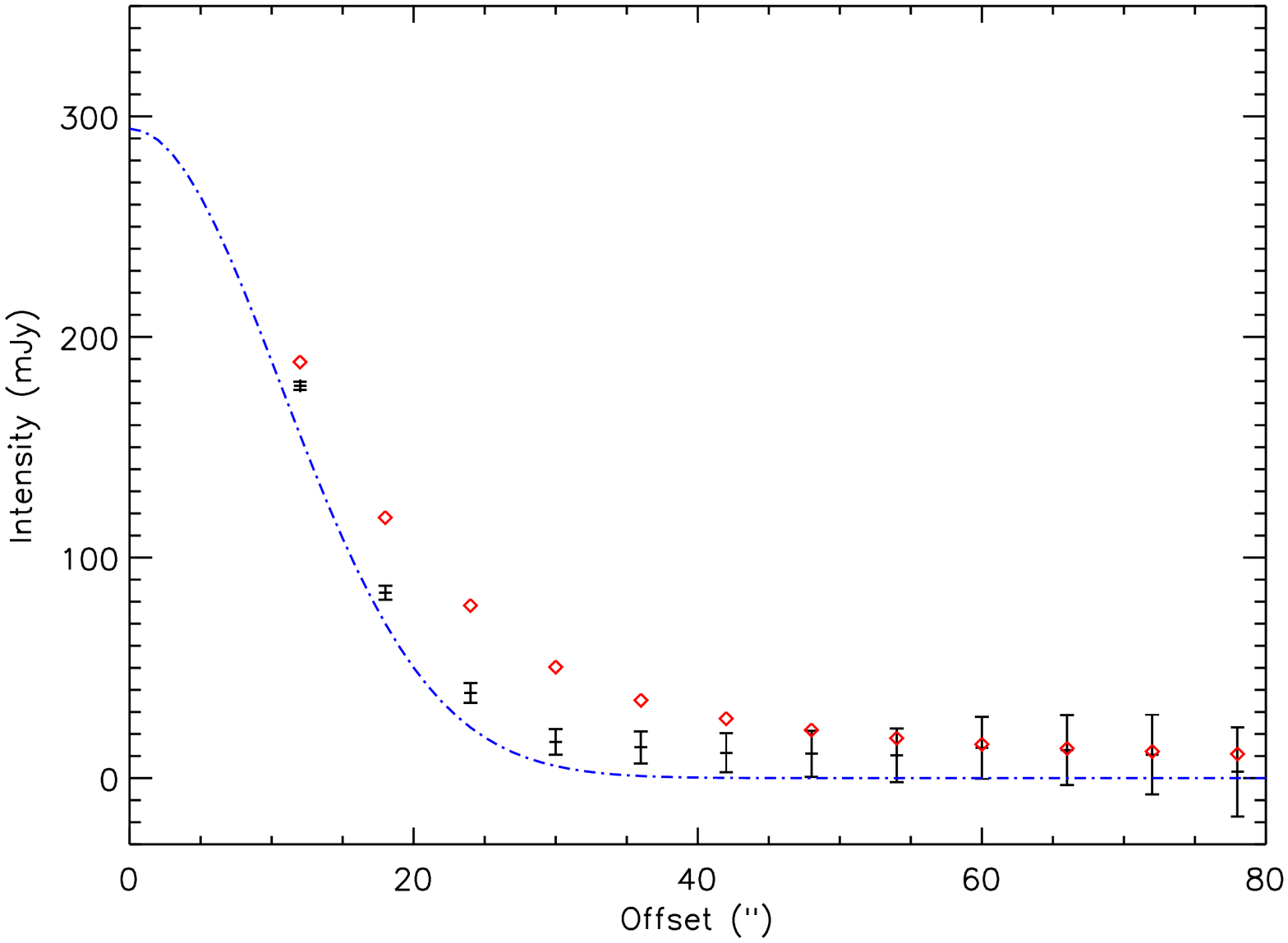}
\end{minipage}
\caption{Left panel: 
IRC\,+10420 ISO-SWS/LWS spectrum (dashed black line) with the best model (continuous red
line) together with the LRS spectrum (purple crosses) with its respective best model (dash-dot blue line). 
The black stars are the photometric data and the red diamonds the LABoCa measurement at 870 $\mu$m.
The inset shows the spectral range between 0 to 50 $\mu$m.
 Right panel:
 Differential aperture photometry on the LABoCa map of IRC\,10420
 (black symbols) compared to the synthetic brightness map derived
 from the best model (red symbols).  
 The dashed blue line is the HPBW.}
\label{fig10420}

\begin{minipage}[c]{.48\linewidth}
\includegraphics[angle=90,width=8.6cm]{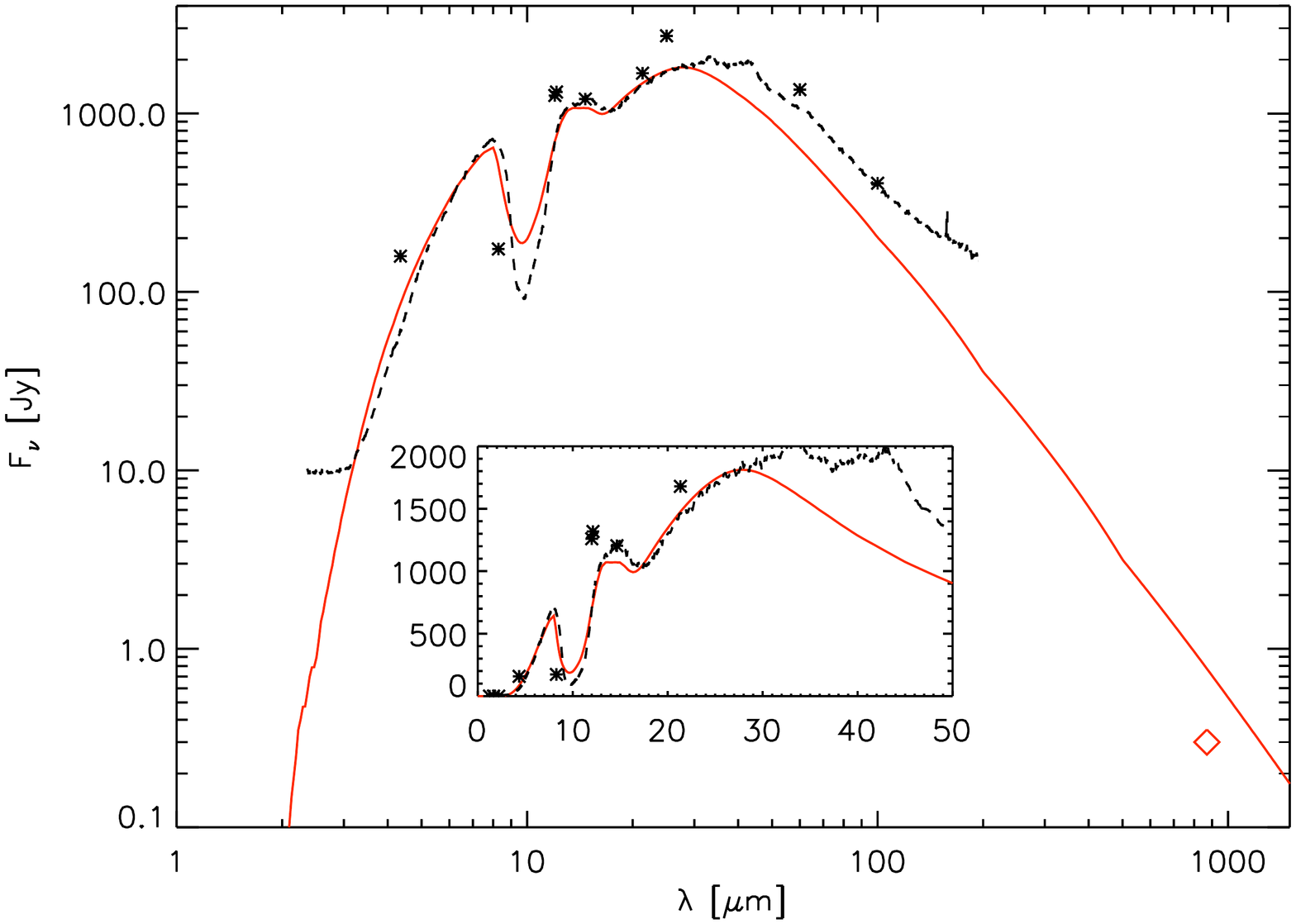}
\end{minipage}
\begin{minipage}[c]{.48\linewidth}
\includegraphics[angle=90,width=8.6cm]{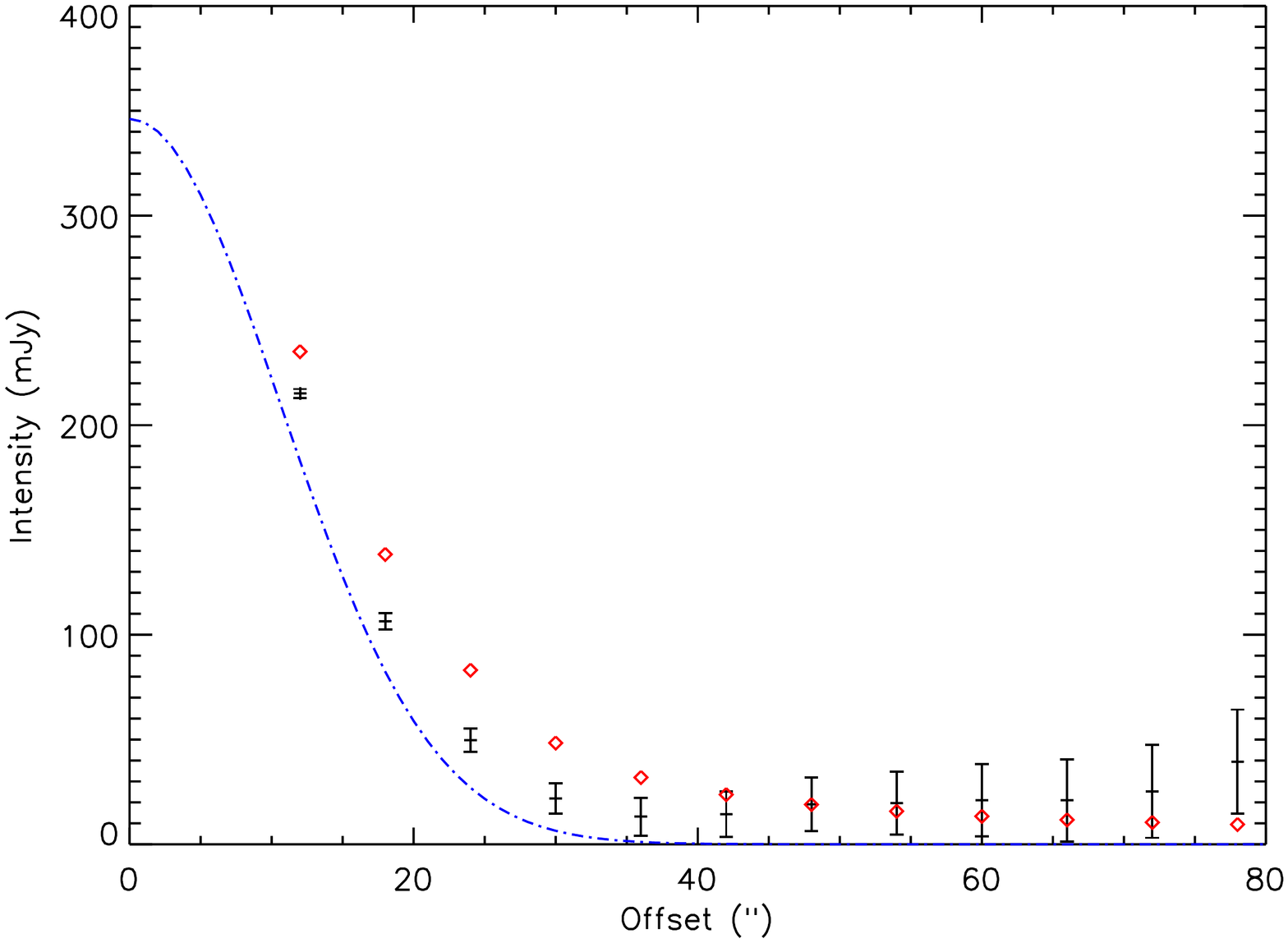}
\end{minipage}
\caption{As Figure~\ref{ocet} for AFGL\,5379.}
\label{fig5379}

\begin{minipage}[c]{.48\linewidth}
\includegraphics[angle=90,width=8.6cm]{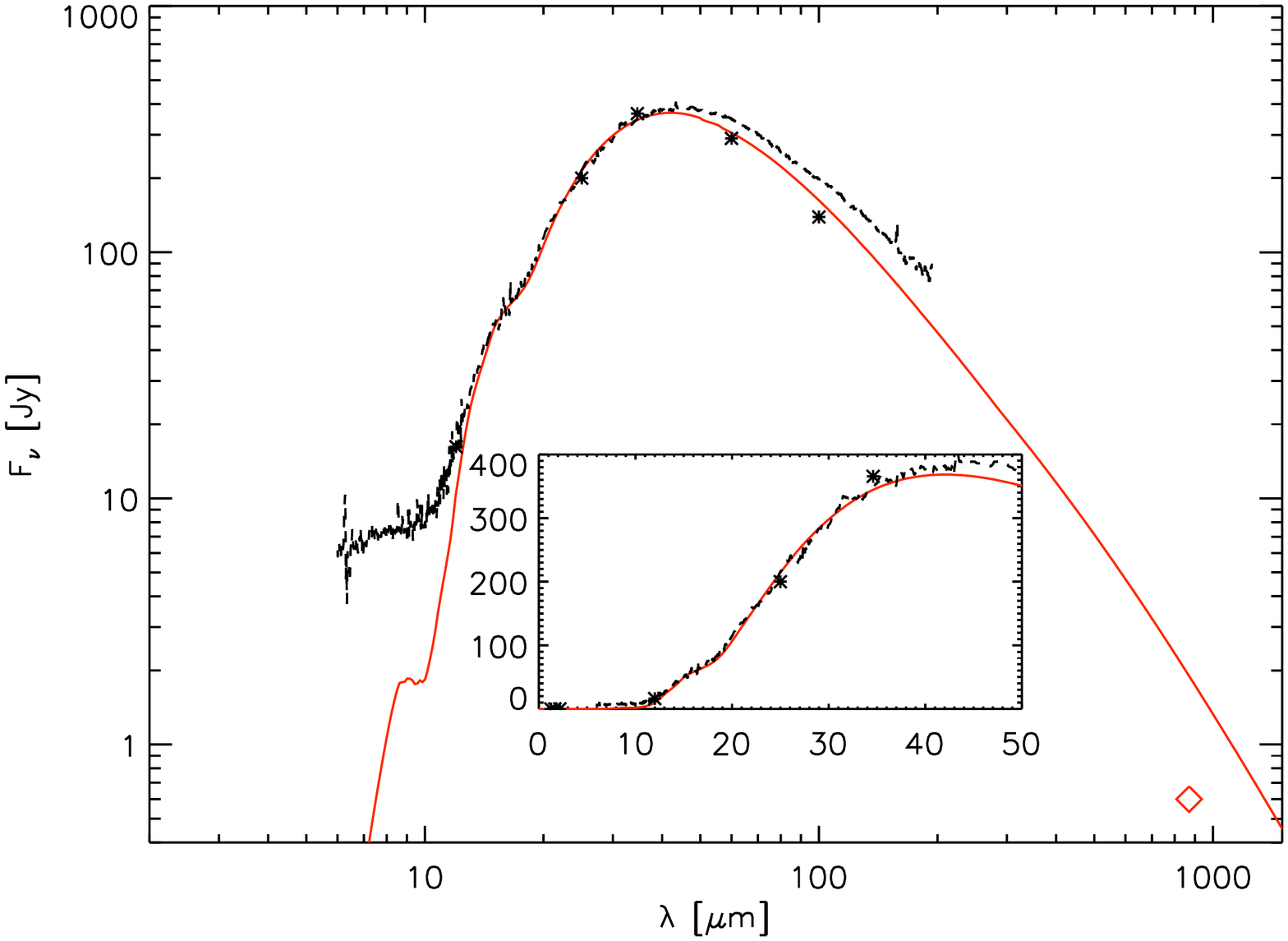}
\end{minipage}
\begin{minipage}[c]{.48\linewidth}
\includegraphics[angle=90,width=8.6cm]{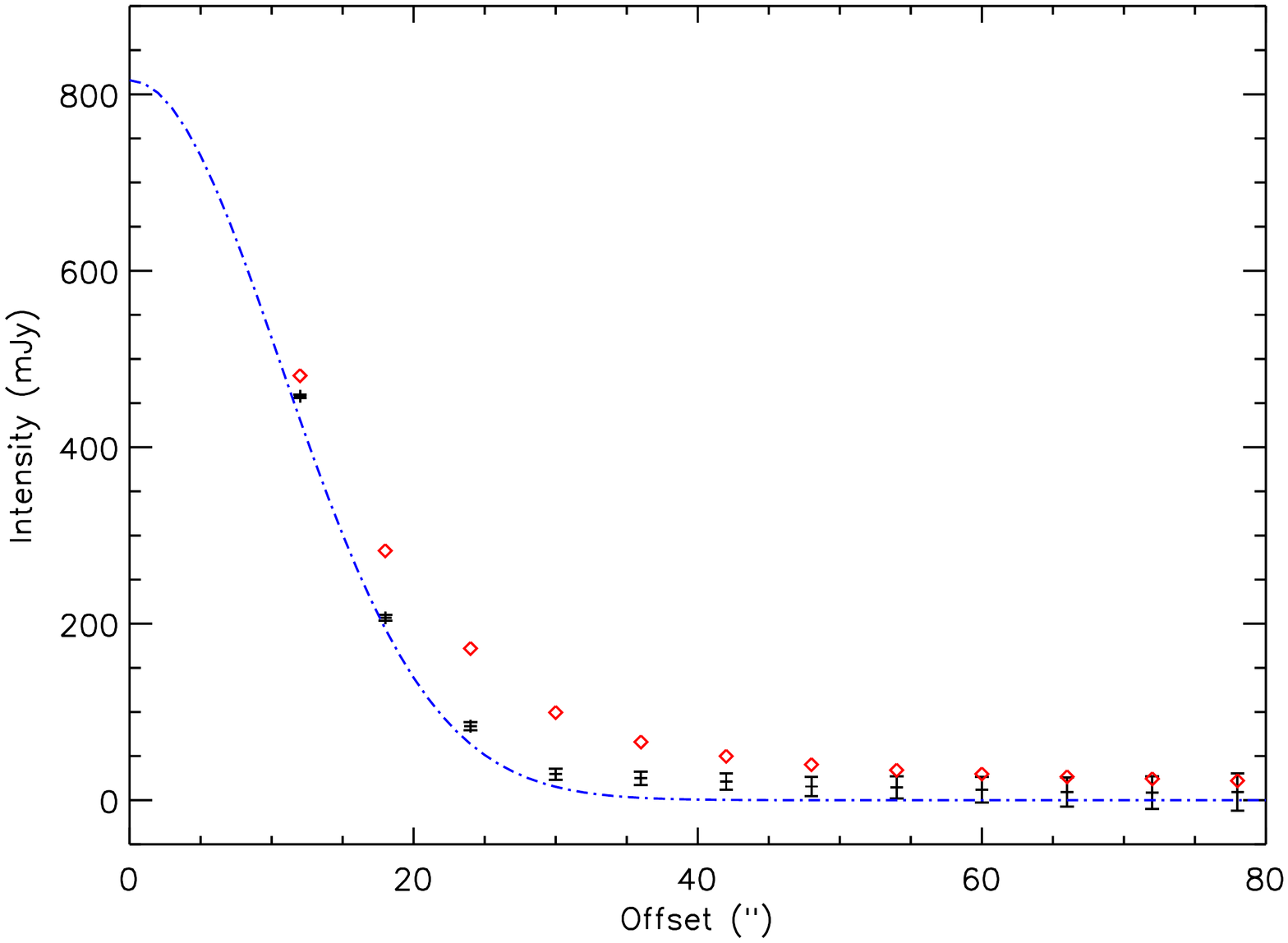}
\end{minipage}
\caption{As Figure~\ref{ocet} for IRAS\,16342$-$3814.}

\label{figLast}
\end{figure}

\end{appendix}

\end{document}